\documentclass[twocolumn,times]{aastex631}

\usepackage{graphicx}
\usepackage{epstopdf}
\usepackage{amssymb}
\usepackage[T1]{fontenc}
\usepackage{gensymb}
\usepackage{rotating}
\usepackage{float}
\def\ltsima{$\; \buildrel < \over \sim \;$}
\def\simlt{\lower.5ex\hbox{\ltsima}}
\def\gtsima{$\; \buildrel > \over \sim \;$}
\def\simgt{\lower.5ex\hbox{\gtsima}}
\newcommand{\msun}{{\rm\,M$_\odot$}}
\newcommand{\sfr}{{\rm\,M$_\odot$\,yr$^{-1}$}}
\newcommand{\lsun}{{\rm\,L$_\odot$}}
\newcommand{\hst}{\textit{HST}}
\newcommand{\spitzer}{\textit{Spitzer}}

\newcommand {\um}{$\mu$m}
\newcommand {\ums}{$\mu$m~}
\newcommand {\ppm}{$\pm$}

\def\um     {$\mu$m}
\def\ts     {\thinspace}
\def\kms    {\ifmmode{{\rm \ts km\ts s}^{-1}}\else{\ts km\ts s$^{-1}$}\fi}
\def\msol   {\ifmmode{{\rm M}_{\odot}}\else{M$_{\odot}$}\fi}
\def\lsol   {\ifmmode{{\rm L}_{\odot}}\else{L$_{\odot}$}\fi}
\def\zsol   {\ifmmode{{\rm Z}_{\odot}}\else{Z$_{\odot}$}\fi}

\def\ltsima{$\; \buildrel < \over \sim \;$}
\def\simlt{\lower.5ex\hbox{\ltsima}}
\def\gtsima{$\; \buildrel > \over \sim \;$}
\def\simgt{\lower.5ex\hbox{\gtsima}}

\shortauthors{Shu et al.}

\begin{document}
\begin{sloppypar}
\correspondingauthor{Xinwen Shu \& Wei-Hao Wang}
\email{xwshu@ahnu.edu.cn, whwang@asiaa.sinica.edu.tw}
\title{A census of optically dark massive galaxies in the early Universe from magnification by lensing galaxy clusters}

\author{Xinwen~Shu}
\affiliation{Department of Physics, Anhui Normal University, Wuhu, Anhui, 241002, China}

\author{Lei~Yang}
\affiliation{Department of Physics, Anhui Normal University, Wuhu, Anhui, 241002, China}

\author{Daizhong~Liu}
\affiliation{
		Max-Planck-Institut f\"ur Extraterrestrische Physik (MPE), Giessenbachstr. 1, D-85748 Garching, Germany
	}

\author{Wei-Hao~Wang}
\affiliation{Academia Sinica Institute of Astronomy and Astrophysics (ASIAA), No. 1, Section 4, Roosevelt Rd., Taipei 10617, Taiwan}

\author{Tao~Wang} 
\affiliation{School of Astronomy and Space Science, Nanjing University, Nanjing
210093, China}

\author{Yunkun~Han} 
\affiliation{ Yunnan Observatories, Chinese Academy of Sciences, 396 Yangfangwang, Guandu District, Kunming, 650216, China }

\author{Xingxing~Huang} 
\affiliation{ CAS Key Laboratory for Researches in Galaxies and Cosmology, Department of Astronomy, University of Science and Technology of China, Hefei, Anhui 230026, China }

\author{Chen-Fatt~Lim} 
\affiliation{Academia Sinica Institute of Astronomy and Astrophysics (ASIAA), No. 1, Section 4, Roosevelt Rd., Taipei 10617, Taiwan}

\author{Yu-Yen~Chang} 
\affiliation{Academia Sinica Institute of Astronomy and Astrophysics (ASIAA), No. 1, Section 4, Roosevelt Rd., Taipei 10617, Taiwan}

\author{Wei~Zheng} 
\affiliation{Department of Physics and Astronomy, Johns Hopkins University, Baltimore, MD 21218, USA}

\author{Xianzhong~Zheng} 
\affiliation{Purple Mountain Observatory, Chinese Academy of Sciences,
	      Nanjing, 210008, China}

\author{Junxian~Wang} 
\affiliation{ CAS Key Laboratory for Researches in Galaxies and Cosmology, Department of Astronomy, University of Science and Technology of China, Hefei, Anhui 230026, China }	      
	      
\author{Xu~Kong} 
\affiliation{ CAS Key Laboratory for Researches in Galaxies and Cosmology, Department of Astronomy, University of Science and Technology of China, Hefei, Anhui 230026, China }



\begin{abstract}
We present ALMA 870\ums and JCMT/SCUBA2 850\ums dust continuum observations 
of a sample of optically dark and strongly lensed galaxies in the cluster fields. 
The ALMA and SCUBA2 observations reach a median rms of $\sim$0.11 mJy and $0.44$ mJy, respectively, 
with the latter close to the confusion limit of the data at 850\um. This represents one of the most sensitive 
searches for dust emission in optically dark galaxies. We detect the dust emission in 12 out of 15 galaxies at $>$3.8$\sigma$, 
corresponding to a detection rate of 80\%. Thanks to the gravitational lensing, we reach a deeper limiting flux  
than previous surveys in blank fields by a factor of $\sim3$. 
We estimate de-lensed infrared luminosities in the range $2.9\times10^{11}-4.9\times10^{12}$\lsun, 
which correspond to dust-obscured star formation rates (SFRs) {of $\sim30-520$\sfr. } 
Stellar population fits to the optical-to-NIR photometric data yield a median redshift $z=4.26$ and 
de-lensed stellar mass $6.0\times10^{10}$\msun. 
{They contribute a lensing-corrected star-formation rate density at least an order of magnitude higher than 
that of equivalently massive UV-selected galaxies at $z>3$. }
The results suggest that there is a missing population of massive star-forming galaxies in the early Universe, which may dominate the 
SFR density at the massive end ($M_{\rm \star}>10^{10.3}$\msun). 
Five optically dark galaxies are located within $r<50$\arcsec~ in one cluster field, 
representing a potential overdensity structure that has a physical origin at a confidence level $>$99.974\%
from Poisson statistics. 
Follow-up spectroscopic observations with ALMA and/or JWST are crucial to confirm whether it is 
associated with a protocluster at similar redshifts. 

\end{abstract}

\keywords{galaxies: evolution -- galaxies: clusters -- galaxies: high-redshift:  -- gravitational lensing -- submillimeter: galaxies}

\section{INTRODUCTION}\label{sec:intro}

Deep surveys with the {\it Hubble Space Telescope} (\hst) and 
the {\it Spitzer Space Telescope} (\spitzer) have greatly enriched our knowledge of the 
early formation history of galaxies. 
By pushing \hst~to its limits, deep imaging of both blank and  
lensing galaxy cluster fields have yielded high-redshift candidates out to $z\sim10$ 
\citep[e.g.,][]{Zheng2012, Coe2013, Oesch2013, Bouwens2014, Salmon2018}. 
Most of these high-redshift galaxies are selected with the Lyman break "dropout" technique 
which relies on the strong break in the rest-frame ultraviolet (UV) spectrum of young 
star-forming galaxies \citep[][and references therein]{Steidel1995}. 
Although it has been proven as one of the most effective ways to find high-redshift galaxies, 
the Lyman break selection is biased towards young star-forming galaxies with little 
dust attenuation, leading to a significant number of {\it dusty} star-forming galaxies at high redshift 
missed by rest-frame ultraviolet (UV) surveys even with \hst~\citep[e.g.,][]{Wang2009, Chen2014, Wang2016, Wang2019, Franco2018, Smail2021}. 

The discovery of (sub)millimeter-bright, dusty star-forming galaxies at high redshifts 
has revolutionized our understanding of the star formation in the early Universe \citep{Chapman2005, Casey2014, Smocic2015, Liu2018, Stach2019},  
specifically those obscured star-forming activities, providing 
more complete census of star-formation history over cosmic times. 
Such submillimeter galaxies (SMGs) are heavily attenuated in the rest-frame UV \citep{Simpson2014, Casey2014}, and therefore are generally 
missed in studies of Lyman break galaxies (LBGs). 
At $z\simgt4$, most SMGs show no counterparts at optical, and in some cases even 
not been detected in the deep near-IR imaging. For instance, 
\citet{Wang2009} performed ultra-deep \hst~ $H$-band observations of an  
SMG in the GOODS-North field (GN10) down to $\sim$29.5 mag, yet no  
counterpart was found. 
This galaxy is instead detected at mid-IR wavelengths $\lambda\simgt3.6$\um~with the \spitzer/IRAC 
($>1\mu$Jy), yielding an extremely red color of $H$-[3.6\um]$>$4.0. 
Thanks to the sensitive observations of CO emission lines, GN10 has now been 
spectroscopically confirmed to be at $z=5.303$ \citep{Riechers2020}. 

However, the sensitivity limits of most current submillimeter surveys only allow for the study of the most extremely
star-bursting systems \citep[e.g.,][]{Asboth2016, Geach2017, Simpson2019}, which may represent merely a tip of ice-berg 
of dust-obscured star formation {in the early Universe.} A potential population of more typical dusty star-forming galaxies at high-$z$ 
is still to be found \citep{Wang2017}. 
The Atacama Large Millimetre/submillimetre Array (ALMA) has now opened a new avenue to
refine our understanding of dusty galaxies at high redshifts, enabling to uncover 
faint SMGs down to a flux level of $0.1-1$ mJy. 
Several ALMA blind surveys have been performed and allowed to detect and characterize the faint SMGs 
{across the cosmic time \citep[e.g.,][]{Aravena2016, Wang2016, Dunlop2017, Umehata2017, Franco2018, Yamaguchi2019, Gonzalez-Lopez2020}.  }
Based on the ALMA survey of GOODS-South field over an area of 69 arcmin$^2$, \citet{Franco2018} found 
that 20\% of the 1.1mm sources are not detected with \hst~ down to a depth of $H\sim28$ mag, and 
suggested that they are massive 
main-sequence star-forming galaxies at $z>4$ \citep[see also,][]{Yamaguchi2019, Umehata2020}. 
A similar fraction of {\it HST}-dark galaxies has also been found in the ALMA {\sc [C ii]} survey of main-sequence 
galaxies at $4.4<z<5.9$ \citep[$\sim$14\%,][]{Gruppioni2020}. 
Conversely, the existence of such \hst-dark galaxies can also be uncovered by focusing the reddest galaxies in 
the IRAC and $H$ bands ($H$-[4.5$\mu$m]$>$4.0), namely $H$-dropouts \citep{Huang2011, Caputi2012, Wang2016}. 
Follow-up continuum observations with ALMA of a sample of 63 $H$-dropouts have yielded detections of 39 
sources down to an 870\ums flux density of 0.6 mJy \citep{Wang2019}. 
They further suggested that the ALMA-detected $H$-dropouts are the bulk populations of massive 
($M_{\star}\simgt10^{10.3}$\msun) star-forming galaxies at $z>3$ with the contribution to 
star-formation rate density an order of magnitude higher than that of equivalently massive LBGs. 
To uncover the nature of $H$-dropouts and the critical role they play in the cosmic evolution of massive star-forming 
	galaxies, we need to explore the fainter population that might have even fainter (sub)millimeter fluxes. 

Complementary to these studies focusing on $H$-dropout galaxies (optically dark galaxies, hereafter)
\footnote{Although some galaxies that are not detectable at H-band in public catalog, 
there is weak residual emission recored if measured using the positional priors from ALMA \citep[e.g.,][]{Zhou2020}. 
Therefore we refer them as optically dark galaxies throughout the paper. }
in blank fields are the searches behind strongly lensing galaxy clusters. 
Gravitational lensing enables the 
detection of intrinsically fainter distant galaxies 
than those otherwise not accessible with {direct ``blank'' field surveys 
\citep[e.g.,][]{Kneib2004, Bradley2008, Zheng2012, Coe2015, Watson2015, Salmon2020, Fujimoto2021, Heywood2021, Laporte2021}. } 
In this paper, we present a systematic search for optically dark galaxies behind 
{31 lensing} clusters using mainly the data from the Cluster Lensing And Supernova survey with
Hubble (CLASH) and Hubble Frontier Fields (HFF). 
Follow-up observations of 15 optically dark galaxies have been performed with 
ALMA and JCMT, aiming at to unveil the population of intrinsically faint dusty star-forming galaxies 
at high-redshifts. 
{This paper is organized as follows. 
We describe the observations, data and selection of optically dark galaxies in Section 2. 
The lensing models, photo-$z$ analysis and Spectral Energy Distribution (SED) fittings are described in Section 3. 
In Section 4, we present the results and discussions, which are summarized in Section 5.  
All magnitudes are in the AB system, and the stellar initial mass function of \citet{Chabrier2003} is assumed 
throughout. 
We adopt the cosmological parameters ($\Omega_{\rm M}$, $\Omega_{\Lambda}$, $h$)= (0.30, 0.70, 0.70). 
}


\begin{figure*}[htbp]
\centering
\includegraphics[scale=0.5]{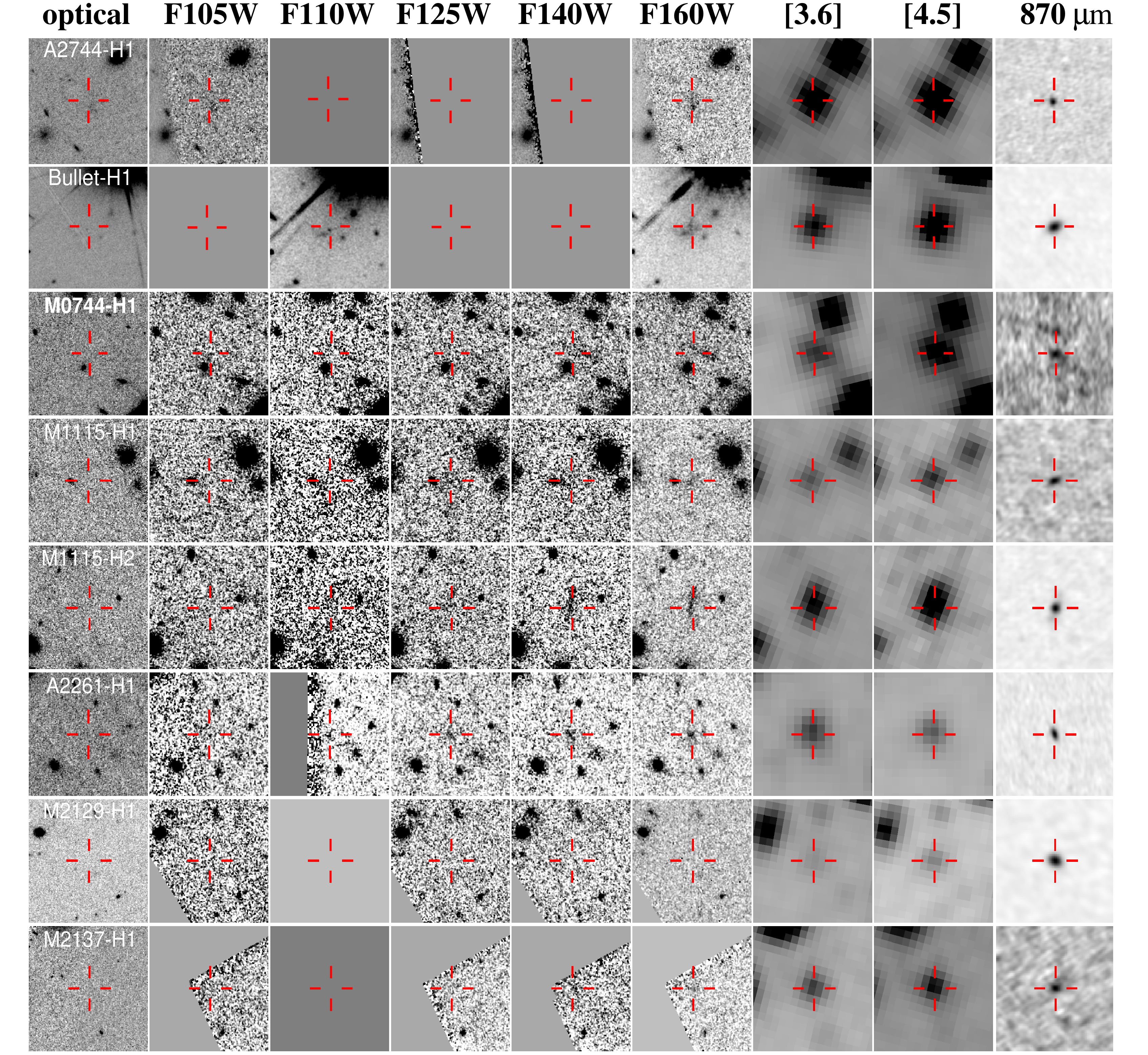}
\label{image}%
\caption{
Postage-stamp images of eight optically dark galaxies detected by ALMA. 
From left to right, the optical images from the respective ACS detection images, which are the 
weighted sums of ACS data at all optical bands available, WFC3/NIR images in the F105W, F110W, F125W, F140W, and F160W bands, 
IRAC 3.6\ums and 4.5\ums images, and ALMA 870\ums images. Each panel has a size of 10\arcsec$\times$10\arcsec. 
{The empty regions indicate either the source is out of the field of view, or not observed in the given {\it HST}/WFC3 filters. }
 }
\end{figure*}

\section{Observations and Data}
\label{sec:obs}
\subsection{HST data}
CLASH is a 524-orbit HST multi-cycle treasury program to observe 25 galaxy
clusters \citep{Postman2012}.  Each cluster was observed using WFC3/UVIS,
ACS/WFC and WFC3/IR to obtain images in 16 broadband
filters spanning from $0.2 - 1.7$$\mu$m. 
The nominal limiting magnitudes in the \hst~WFC3/IR bands 
are approximately 27.2-27.8 (5$\sigma$ detection limit in a 0.4\arcsec~diameter aperture). 
The CLASH data were processed based on the {\it Mosaicdrizzle} pipeline \citep{Koekemoer2011}, 
which is described in detail in \citet{Postman2012}. 
The pipeline produces cosmic-ray rejected and aligned images for each filter with 
a common pixel scale of 0$\farcs$065. 
The multi-wavelength photometric catalogs were extracted using the software {\tt SEXTRACTOR} \citep{Bertin1996}, 
and made retrievable in the CLASH data archive\footnote{http://archive.stsci.edu/prepds/clash/}.   

The HFF program carried out deep \hst~imaging of six clusters as well as six parallel fields, 
with four WFC/IR filters (F160W, F140W, F125W and F105W), 
and three ACS filters (F435W, F606W, F814W). 
Totally 840 \hst~orbits have been allocated to HFF, resulting in a survey depth of $m_{\rm AB}$$\sim$29 mag. 
Among six HFF clusters, four have previously observed as part of the CLASH survey 
(MACS0416.1-2403, MACS0717.5+3745, MACS1149.6+2223 and Abell 1063S). 
Details of the HFF survey design are provided in \citet{Lotz2017}. The data were reduced by the HFF team using the 
same pipeline as that to produce the CLASH images, 
which are available at the HFF data archive\footnote{https://archive.stsci.edu/prepds/frontier/}. 
In addition, we also utilized the data taken from the \hst~observations of four other clusters, 
{including Bullet, Abell 1689, Abell 1703 and Abell 2218, which have been extensively studied in 
searching for high-redshift lensed LBGs \citep{Bradley2008, Bradley2012, Hall2012}. 
Their WFC3/IR data reach 5$\sigma$ limiting magnitudes of $26.9-27.3$ mag, 
similar to that of CLASH.  }  
Totally 31 cluster lensing fields have been compiled in this paper. 

\subsection{Spitzer data}

Deep \spitzer/IRAC observations of CLASH cluster fields 
in the Channels 1 ($\sim$3.6\ums) and 2 ($\sim$4.5\um) were provided 
by the ICLASH (GO 80168; PI: Bouwens), the Ultra-Deep IRAC imaging 
of Massive Lensing Galaxy Clusters (GO 20439; PI Egami), 
the \spitzer~IRAC Lensing Survey program (GO 60034; PI: Egami),  
and \spitzer~UltRa Faint SUrvey Program \citep{Bradac2014}. 
The typical exposure for CLASH cluster is 3.5-5 hr per IRAC band, reaching 
3$\sigma$ sensitivity limits of $\sim$26 AB magnitude.  
As part of the HFF campaign, deep \spitzer/IRAC images were obtained down to 
depths of 26.5 and 26 mag (5$\sigma$) in the six cluster and parallel fields, respectively. 
For each cluster, we used the deepest \spitzer/IRAC observations when available, and 
processed the IRAC corrected Basic Calibrated Data (cBCD) images with {\tt MOPEX}, following 
the procedures described in \citet{Zheng2014}. The images were sampled to a final pixel scale of 0$\farcs$6. 

\subsection{Selection of optically dark galaxies}

In order to select magnified massive star-forming galaxies that are optically dark,  
we first cross-matched the \hst~WFC3/IR selected catalogue\footnote{The catalog was produced by combining 
the images from five/four WFC3/IR filters for CLASH/HFF as the detection image in {\tt SEXTRACTOR}.} 
with the IRAC 4.5\um-selected catalogue for each cluster field, and identified those IRAC sources without WFC3/IR counterparts 
within a matching radius of $r=2$\arcsec~(corresponding roughly to the point-source FWHM of IRAC/4.5\um~observations). 
IRAC sources have been inspected visually to ensure that they are not artifacts such as hot pixels, 
diffraction spikes of stars, or extended emission from brighter foreground galaxies. 
{This leaves totally nine 4.5\um~sources that can be classified as $H$-band ``dropouts" at the depth of the CLASH and HFF}.
{We also take into account eight optically dark galaxies }with extremely red colors of $H$-[4.5\um]$\simgt$4, as they  
likely represent similar massive star-forming galaxies at $z>3$ \citep{Huang2011, Caputi2012, Wang2016}. 
Without lensing magnification, they would be equivalently selected as $H$-band ``dropouts" 
at the depth of blank field surveys. 
The final sample consists of total 17 optically dark galaxies, 
among which 10 are observed with ALMA and five observed 
with SCUBA2 (Section 2.4). 
Note that these extremely red objects are not distributed uniformly across clusters, 
possibly due to the cosmic variance. 
Field M0429 appears unique among the survey with five objects satisfying  
the selection criteria for optically dark sources, 
which will be investigated in detail in Section 4.3. 


\subsection{Submillimeter observations}
Our ALMA Band-7 continuum observations of ten optically dark galaxies were performed during January and July 2018 
(project 2018.1.01409.S, PI: Wang). 
The observations were centered on the IRAC positions with a spectral configuration covering 
a $\sim$7.5 GHz bandwidth and a central frequency of around 343.5 GHz (870\um). 
The on-source integration time is roughly 5--10 min per source. 
We reduced the data using the CASA software \citep[version 4.3.1]{McMullin2007}, 
following the standard calibration pipeline provided by the observatory. 
The calibrated data were imaged using the {\tt tclean} task of CASA, resulting in a typical 
synthesized beam of 0$\farcs$3-0$\farcs$6 and rms of $\sim$$50-200 \mu$Jy/beam. 
We measured the total flux of all our targets directly in the (u,v) plane using
the {\tt uvmodelfit} task of CASA, assuming a Gaussian model. 
Eight out of ten targets were detected at S/N$>$4 with $S_{870\mu m}\simgt$0.6 mJy. 
The positions of the 870\ums emission as measured from
ALMA are in good agreement with the IRAC ones, with a median positional 
offset 0$\farcs$2 ($<$0$\farcs$6).  

In addition to ALMA observations, we also observed the cluster field M0429 at 850\ums 
with JCMT/SCUBA2 (Program ID: M16BP006), 
which shows a possible overdensity structure of optically dark galaxies. 
The JCMT observations were carried out using the CV {\sc DAISY} scan pattern which provides 
a 12\arcmin~diameter map with uniform exposure coverage in the central 3\arcmin~ diameter region. 
In total 27.4 hours of observations were acquired between 2016 Aug and 2016 Nov, consisting of 
nine separate $\sim$1.1-4.5 hours long scans at weather conditions of $\tau=[0.08, 0.12]$. 
We reduced the data using the Dynamic Iterative Map Maker (DIMM) within the STARLINK sub-Millimeter User 
Reduction Facility (SMURF) software package \citep{Chapin2013}, and 
calibrated the data with the flux conversion factors of 537 Jy pW$^{-1}$ beam$^{-1}$.  
The SCUBA-2 850\ums maps for each individual exposure were co-added and combined, using the STARLINK PIpeline for Combining 
and Analyzing Reduced Data (PICARD), and beam-match filtered with a 15\arcsec~FWHM Gaussian kernel. 
{Although the spatial resolution for JCMT/SCUBA2 observations is poor ($\sim$14\arcsec),} the final map reaches a rms of 
0.44 mJy/beam in the central region, close to the confusion limit of the data at 850\um. 

\section{Models}
\subsection{Gravitational Lensing magnification}
In addition to the uniform multi-wavelength photometry provided by \hst, 
another major advantage of CLASH observations is that cluster lensing enables us to discover
intrinsically faint galaxies. The magnification maps for all
25 CLASH clusters were made based on the strong lensing model of
\citet[][]{Zitrin2015}. 
{Two lensing models are available in the archive of CLASH data release. 
One is constructed with the light-traces-mass (LTM) method, which assumes that the mass distributions of 
both the galaxies and dark matter (DM) are reasonably traced by the cluster's light distribution. 
Another model adopts the LTM assumption only for the galaxy component, 
whereas the DM component is modeled as an elliptical Navarro-Frenk-White (eNFW) mass-density distribution. 
Cluster galaxies are each modeled as a pseudo-isothermal elliptical mass distribution, scaled by its luminosity. 
The relative systematic difference is $\sim$20\% in the magnification ($\mu$) 
between the LTM and eNFW models. 
Since both models are equally valuable and useful, as discussed in \citet[][]{Zitrin2015}, 
in the following analysis we adopted the mean magnification from the two models as our fiducial 
estimate of magnification factor for each galaxy. }

As part of the HFF program, seven independent derived lensing models 
of six clusters were developed, and publicly released through the HFF 
data archive. 
{Only one galaxy in our sample was found in the HFF clusters (A2744-H1), 
and we adopted the median value derived from seven models as its best-fit 
magnification factor.} 
As magnification factors are redshift dependent, we assumed a photometric redshift (Section 3.2) 
to estimate lensing magnification for each our galaxy, which is listed in Table 1. 
{Note that there are no publicly available lensing models for the Bullet cluster, 
we conservatively estimated the magnification factor of $\mu>8$ for Bullet-H1, as 
it is only $\sim$15\arcsec~away from the critical curve from the $z = 3-7$ magnification map \citep{Gonzalez2009, Hall2012}. 
In addition, Bullet-H1 is located at a region for which the magnification factors are estimated in the range $\mu=8-34$ 
for the lensed galaxies at $z>6$ \citep{Bradac2009}. 
}
The estimated magnifications are likely consistent with the
true value at 68\% confidence at the low magnification end \citep[e.g., $\mu<$5,][]{Bradley2014}. 
The median magnification from the models is $\mu\sim1.91$, over an average area of 4.5 arcmin$^2$ per cluster. 
{Regardless of Bullet-H1, M0429-H3 has the largest magnification factor of $\mu=7.86$ among 11 submm-detected sources, 
resulting in a delensed flux at 850\ums as low as 0.2 mJy. } 
We will discuss in Section 4.1 how the intrinsic properties of the 
{optically dark galaxies} are revealed with the help of gravitational lensing.

\subsection{Photometric redshifts and stellar properties}

In Section 2.3, we have shown that in our sample there are seven optically dark galaxies 
which have no WFC3/IR counterparts in the public photometric catalog, 
{but detected at IRAC 3.6\um~and 4.5\um.}
With prior knowledge of their positions in the ALMA data, 
we performed manual photometry to determine whether there
is any low-level flux recorded in HST (from 0.2\ums to 1.7\um), using a circular aperture of radius 0$\farcs$6. 
The background and flux errors were estimated in an annulus area with inner and outer 
radius of 1$\farcs$5 and 3\arcsec, respectively. 
We found marginal NIR detections for six galaxies, which are 25.7-26.7 mag in the $H$-band. 
The non-detections ($<1\sigma$) in most other \hst~optical wavelengths 
secure their extreme faintness from optical to NIR. 
The IRAC photometry at 3.6\ums and 4.5\ums for all galaxies was 
derived by using {\tt GALFIT} software \citep{Peng2010}, with the same procedures 
as described in \citet[][]{Zheng2014}. 
{\tt GALFIT} has the advantage to deblend and subtract flux from nearby 
contaminating sources when present, allowing for more accurate flux measurements 
for the source of interest \citep[][]{Zheng2014}. 
{This is important for the IRAC photometry of six sources in the sample, 
which have a close neighbor within a distance of 2\arcsec. 
The IRAC 3.6\ums and 4.5\ums fluxes derived from the {\tt GALFIT} fittings are shown in Table 1.}  
Note that for M0744-H2, though it was not shown in the CLASH public photometric catalog, 
we found marginal detection of $26.38\pm0.41$ mag in the $H$-band, and the $H-$[4.5\um]=2.2. 
This indicates that its $H-$[4.5\um] color is not as red as other galaxies in the sample. 
Since it is not detected by ALMA, we are not including the source in our following statistical 
analysis. 



\begin{figure}[htbp]
\centering
\includegraphics[scale=0.25]{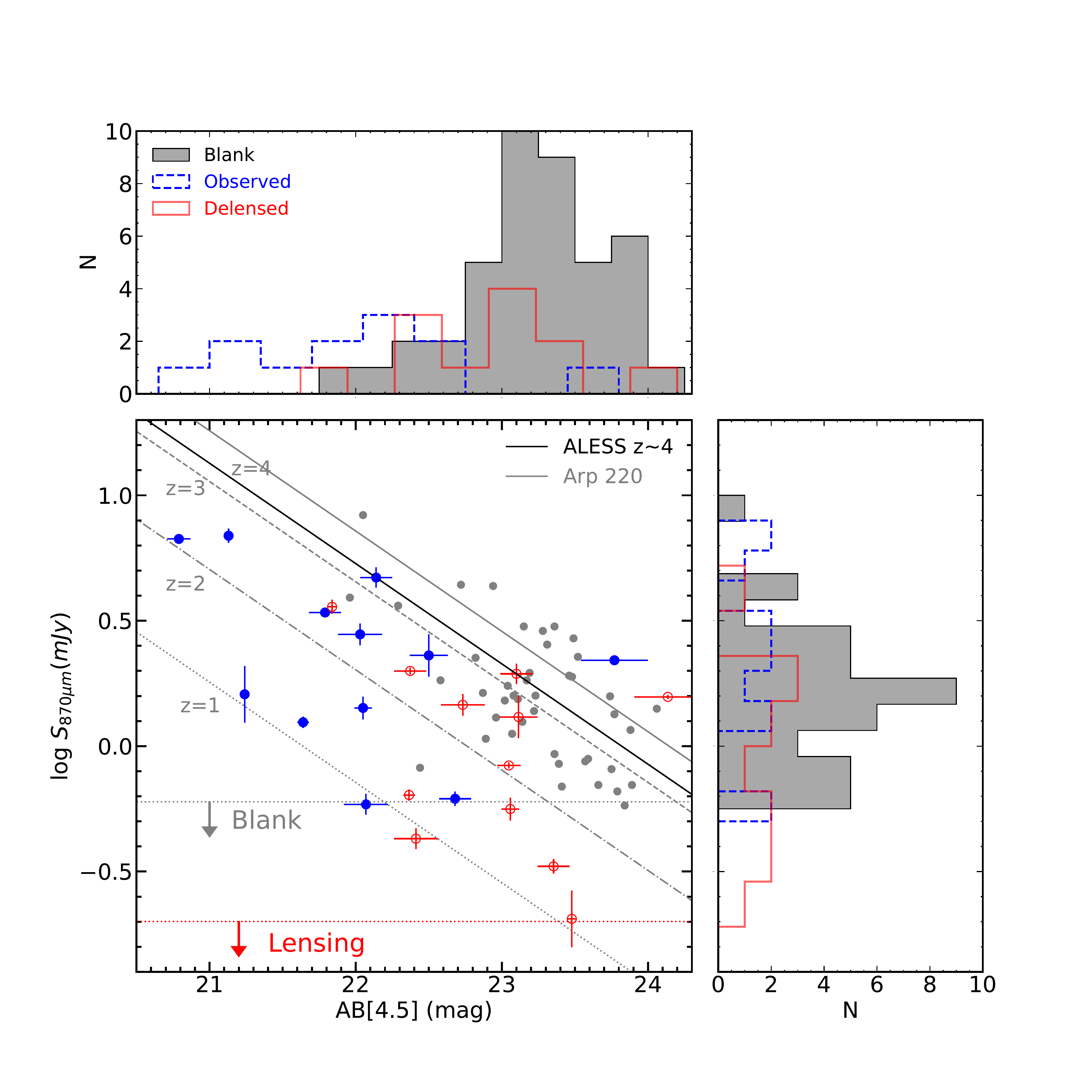}
\caption{The observed IRAC 4.5\ums magnitudes versus 870\ums fluxes for optically dark galaxies 
in the cluster fields (filled blue circles). 
Only the sources detected by ALMA/SCUBA2 are shown for clarity. The de-lensed, intrinsic magnitudes 
and fluxes are shown in open red circles. As a comparison, we also show the sample of 
optically dark galaxies in the blank fields (filler grey circles). 
{The predictions based on the SED templates of ALESS SMGs at $z\sim4$ and Arp 220 at 
various redshifts are plotted in black and grey curves, respectively. }
The horizontal grey dotted line indicates the detection limit in blank field. 
Although the depth of ALMA observations of cluster fields is similar to that of blank field, 
gravitationally lensing magnification allows for detecting fainter fluxes by a factor of three 
(red dotted line). 
The histogram in top panel shows the corresponding distribution of the 4.5\ums magnitudes, while the 
distribution of the 870\ums fluxes is shown in the right panel.  
}
\label{magflux}%
\end{figure}

We calculated photometric redshifts by fitting the multi-wavelength photometry between UVIS/0.2\ums and IRAC/4.5\ums (including 
1$\sigma$ upper limits) of each object with a linear combination of seven galaxy templates, using 
the EAZY photo-$z$ code \citep[][]{Brammer2008}. 
As described in \citet[][]{Brammer2008}, the template set consists of five output templates derived from a 
library of P$\acute{\rm E}$GASE stellar population syntesis models, 
with ages between 1 Myr and 20 Gyr and a variety of star formation histories. 
Additional reddening [$0.5\leq E(B-V)\leq 1.1$] is applied to represent young, dusty 
objects. 
We assigned zero fluxes to the non-detections and corresponding $1\sigma$ errors as the 
flux uncertainties.   
The best-fit templates from EAZY SED fittings suggest a redshift distribution of optically dark 
galaxies in the range $z_{\rm phot}=3.44-5.19$, with a median redshift of $z_{\rm median}=4.26$. 
{For a consistency check, we also used the Bayesian SED modeling and fitting code-BayeSED \citep{Han2012, Han2014, Han2019} 
	to perform SED fittings to the UVIS/0.2\ums to IRAC/4.5\ums photometry. 
	This yields a consistent result of the photo-$z$ distribution within errors (Figure 10 (a) in Appendix B). 
Note that if applying the $L_{\rm IR}$ constraints to the SED fittings, we obtained a slightly 
lower photometric redshifts for the sample, with $z_{\rm median}=3.08$. 
}

\begin{figure}[htbp]
\centering
\includegraphics[scale=0.55]{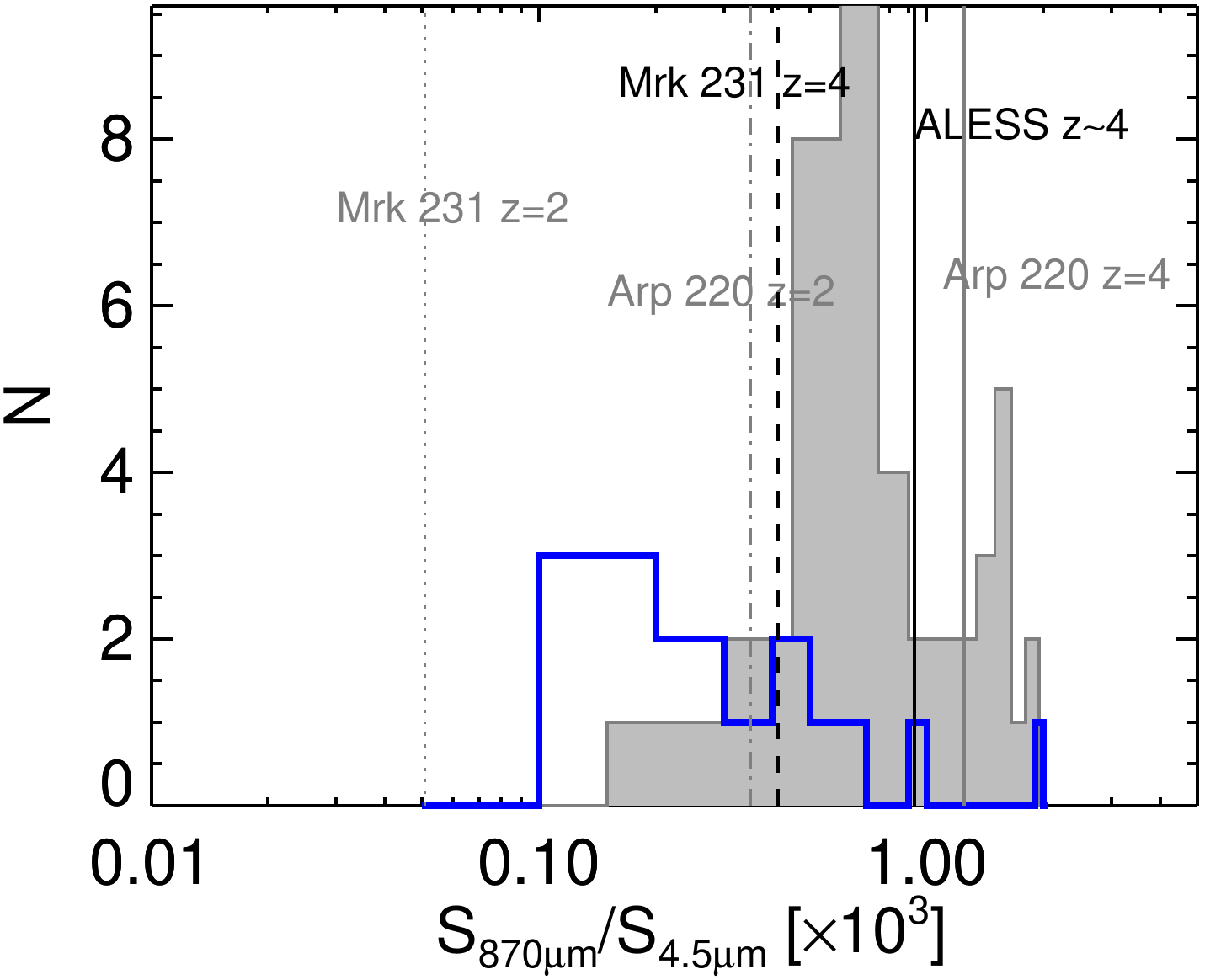}
\caption{
Histogram of the 870\ums to 4.5\ums flux ratios for optically dark galaxies 
in the cluster fields (blue line) and blank fields (grey shaded), respectively. 
The solid vertical line indicates the predicted ratio of $S_{870\mu m}/S_{4.5\mu m}$ based on the 
mean SED of ALESS SMGs at $z\sim4$, and the Arp 220 SED at $z=4$, respectively.  
The dot-dashed line is the predicted flux ratio assuming an Arp 220 SED at $z=2$. 
{The dashed and dotted line represents the predicted flux ratio assuming the SED of dusty 
AGN (Mrk 231) at $z=4$ and $z=2$, respectively. }
}
\label{magratio}%
\end{figure}

We then used FAST code \citep[][]{Kriek2009} to estimate the stellar properties such as stellar masses, 
ages and dust extinctions.  
The \citet[][]{Bruzual2003} stellar population synthesis models with a Chabrier initial mass function and solar 
metallicity are used in FAST. 
We assumed exponentially declining star-formation histories 
with e-folding times in the range $\tau=0.1-10$ Gyr, and adopted the standard
\cite{Calzetti2000} attenuation law with a wide range of dust extinction of $0<A_{\rm V}<8$. 
As another prior, the age of a galaxy is restricted to be less than the age
of the universe at that redshift. 
As shown in Figure 9 of Appendix, the multi-wavelength photometry between UVIS/0.2\ums and IRAC/4.5\ums 
show a clear spectral break between 1.6 and 3.6\ums for most optically dark galaxies. 
The SED fittings suggest that their red $H-[4.5\mu m]$ colors are mainly due to the
redshifted 4000\AA~Balmer break at $z>3$, which is a signature of existence of old 
and evolved stellar population.  
We note that the optical/NIR and IRAC data cannot constrain the stellar age well because of its degeneracy with
the dust attenuation. 
The best-fit SED models from FAST yield delensed stellar masses in the range $M_{\star}=1.0\times10^{10}-9.7\times10^{11}$\msun 
with a median stellar mass of $6.0\times10^{10}$\msun. 
The stellar masses are similar to those in typical SMGs at $z\simgt4$ \citep[e.g.,][]{Michalowski2010, Smocic2015}, 
but are statistically larger than that of UV-selected star-forming galaxies at similar redshifts \citep[e.g.,][]{Stark2009}. 
{Note that the effect on the stellar masses is minor if using the BayeSED to perform the SED fittings by applying the $L_{\rm IR}$ constraints. 
As discussed in detail in the Appendix B, since the dust continuum emission may not completely 
correspond to the one obscuring the stellar component in some high-$z$ galaxies observed by 
ALMA \citep[e.g.,][]{Elbaz2018, Franco2020}, 
we report here only the results of photo-$z$ analysis and SED fittings without applying the $L_{\rm IR}$ constraints. 
}


\section{Results and Discussion}

As shown in Table 1, we detected the submm emission in eight out of ten optically dark galaxies observed with ALMA, 
with a flux down to $\sim$0.6 mJy, i.e., a detection rate of 80\%. The detection rate in the sample of five galaxies 
observed with SCUBA2 is similar. Figure \ref{image} displays the postage-stamp ACS/optical images, WFC3/NIR images, 
IRAC 3.6\ums and 4.5\ums image, and ALMA 870\ums image for eight galaxies detected by ALMA. 
The JCMT/SCUBA2 850\ums imaging of the optical dark galaxies in M0429 field is shown in Figure \ref{m0429}.

\subsection{Massive, dusty star-forming galaxies at $z>3$}

At $z\simgt4$, observed 4.5\um~ fluxes trace the rest-frame optical emission hence the stellar masses \citep[][]{Overzier2009} 
while observed 870\ums fluxes can approximate the FIR luminosities hence obscured SFRs \citep[][]{Cowie2017}.  
We plot the distribution of 870\ums flux versus 4.5\ums magnitude for the lensed optically dark galaxies in Figure \ref{magflux}. 
We compare this distribution with the $H$-dropouts selected in the \hst~CANDELS fields \citep[][]{Wang2019}, 
which are referred as blank (unlensed) fields where no strong lensing effects are expected. 
{For simplification, we only include the sources detected by ALMA (and SCUBA2). 
Our sample is apparently brighter in 4.5\ums than the blank field galaxies.}
To a limiting 4.5\ums magnitude of 22, only one optically dark galaxy is found from unlensed CANDELS fields. 
However, we find five such bright galaxies behind lensing clusters. 
{This effectively illustrates how the gravitational lensing enables to identify}
optically dark galaxies at relatively bright MIR magnitudes, which is helpful for numerous follow-up 
studies, particularly with spectroscopy. 
After correcting for the model magnification factors (see Table 1), a 
Kolmogorov-Smirnov (K-S) test for the 4.5\ums magnitudes 
in lensed and unlensed samples results in a $p$-value of $0.104$, suggesting 
no significant difference between the two distributions.  

\begin{figure}[htbp]
\centering
\includegraphics[scale=0.5, angle=0]{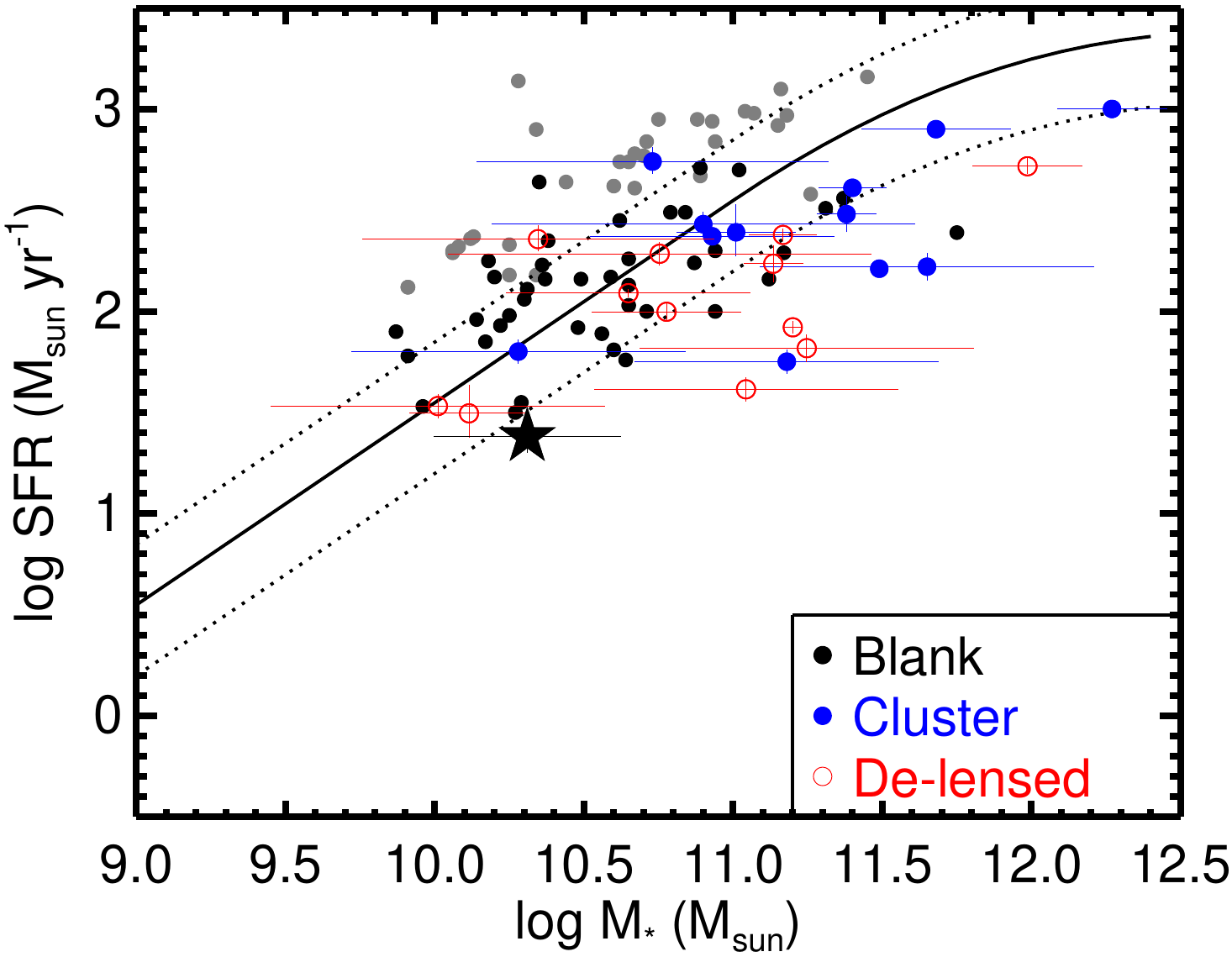}
\caption{
Relation between stellar masses and SFRs for optically dark galaxies in the cluster fields (filled blue circles). 
The relation by corrected for the lensing effects is shown in open red circles. 
For comparison, we also plot the optically dark galaxies in the blank fields (filled black circles), 
and the ALESS SMGs at $z>3.5$ (filled grey circles). The error bars are not shown for clarify. 
The large black star indicates the median stellar mass for the optically dark galaxies in the blank fields 
that are not detected by ALMA, for which the SFR is derived from their stacked ALMA flux.   
}
\label{massvsfr}%
\end{figure}

Figure \ref{magflux} also shows that while the detection limits of ALMA observations are similar between lensing and blank fields, 
i.e., sensitive to sources with flux $S_{\rm 870\mu m}\simgt$0.6 mJy, our sample extends even fainter flux at 870\ums once 
corrected for the gravitational lensing amplifications. 
The faintest flux reaches $\sim$0.2 mJy, which is approximately 3 times lower than 
that of ALMA-detected ones in the blank fields, but consistent with the stacked 
flux of those not detected by ALMA \citep[0.24 mJy,][]{Wang2019}. 
The median delensed 870\ums flux for our sample is 1.2 mJy, which is a factor of 1.3 fainter 
than that of ALMA-detected ones in the blank fields. 
We stress that the uncertainties of strong lensing effect would not affect the comparison, 
as the model lensing magnifications are moderate, with a median value of $\mu=1.91$. 
The results suggest that gravitational lensing allows us to reach deeper limiting fluxes (by a factor of $\sim$3), 
thus revealing the presence of a population of intrinsically faint star-forming galaxies at high redshifts. 
{On the other hand, it can be seen in Figure \ref{magflux} that for a given 4.5\ums magnitude, the lensed ones tend to 
have fainter 870\ums fluxes, suggesting smaller dust-stellar luminosity ratios and therefore less dusty.   
By comparing with the predictions of empirical galaxy SED templates such as Arp 220 for redshifts from 1 to 4, 
Figure \ref{magflux} suggests that lensed ones can have $S_{870\mu m}/S_{4.5\mu m}$ flux ratios consistent with 
that at $z\sim2$. 
In Figure \ref{magratio} we compared the distribution of $S_{870\mu m}/S_{4.5\mu m}$ for lensed galaxies from 
this work and that in blank fields. The median flux ratios for the two samples are $4.88\times10^{2}$ and $7.47\times10^{2}$, 
respectively. A K-S test ($p=0.000486$) suggests that they are not drawn from the same population.  
Note that for a given 870\ums flux, the presence of bright Active Galactic Nucleus (AGN) might boost the 4.5\ums magnitude.  
{Figure \ref{magratio} also displays the predicted flux ratio assuming the SED of dusty AGN, Mrk 231, at 
$z=2$ and $z=4$, respectively, which is indeed smaller than that derived based on the pure galaxy templates. 
Due to the lack of sensitive X-ray observations, we cannot fully rule out the AGN contamination based on the current data, 
so the above comparison should be treated with caution. }
}

\begin{figure}[!htbp]
\centering
\includegraphics[scale=0.52]{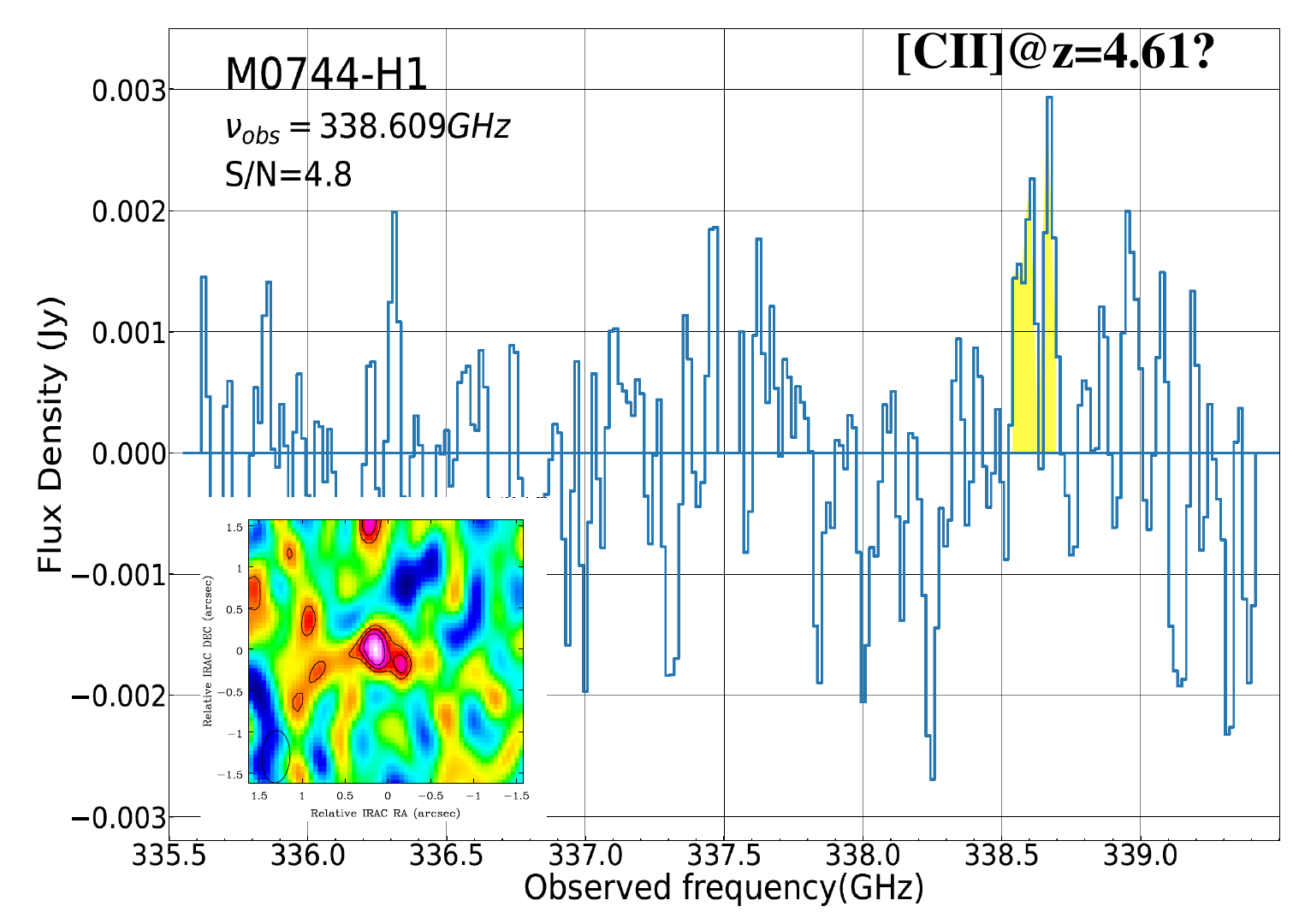}
\caption{
The ALMA Band-7 spectrum in the frequency range 335.5-339.5 GHz for the source M0744-H1. 
The potential emission line detected at $\nu_{\rm obs}=338.609$ GHz is highlighted with yellow 
shaded region. {The continuum has been subtracted to generate the spectrum. 
The velocity-integrated emission line map is shown in the bottom-left panel. 
The black contours are drawn at (3, 4, 5) times the off-source r.m.s. 
The map is centered at the position of the IRAC source at 3.6\um. }
}
\label{cii}%
\end{figure}

In Figure \ref{massvsfr}, we present the relation between the stellar masses and SFRs for our sample. 
As in Figure \ref{magflux}, ALMA-detected optically dark galaxies in the blank fields are 
plotted for comparison. Following \citet[][]{Wang2019}, the SFRs are derived from the 870\ums flux densities by assuming 
that their intrinsic FIR SED resembles that of the stacked one. In addition, we also included the $z>3.5$ SMGs from the ALESS survey \citep[][]{deCunha2015}. 
The solid and dashed lines represent the star-forming main sequence at $z=4$ and its 1$\sigma$ scatter \citep[][]{Schreiber2015}. 
It can be seen that the optically dark galaxies in the lensing cluster fields are characterized by massive dusty star-forming galaxies, which 
{are located around the main sequence at $z=4$, albeit with a larger scatter. After corrected for the lensing effects, 
while most are consistent with the distribution of optically dark galaxies found in the blank fields, 
there are several galaxies (3/12) falling slightly below the star-forming main sequence at $z=4$.  
The result suggests that a fraction of massive, normal star-forming galaxies at $z\sim4$ is indeed missed in previous NIR surveys even with \hst. 
}
It should be noted that approximately 40\% optically dark galaxies in the blank fields are not detected {by ALMA \citep{Wang2019}. }
{In Figure \ref{massvsfr} we show their median SFR derived by a stacking analysis and the 
median stellar mass (filled star), suggesting that they have lower specific SFRs (on average) in comparison with the ALMA-detected ones 
\citep[see also, ][]{Wang2019}. 
The fraction is 20\% for our sample, but we cannot perform a similar statistical analysis as only two galaxies are not detected by ALMA. }
In principle, gravitational lensing would allow to detect 
the dust emission (if presents) from these intrinsically faint galaxies. 
Due to the limited number of galaxies observed by ALMA, our sample does not significantly extend to the region of lower specific SFRs. 
Future ALMA observations of a larger sample of optically dark galaxies in cluster fields will uncover 
even fainter objects. This is important to verify the presence of optically dark galaxies with very low specific SFRs, 
which may represent the rare population of quenched galaxies in the early {Universe \citep[e.g.,][]{Schreiber2018, Mawatari2020, Santini2021}}. 


It is extremely difficult to measure the redshift of optically dark galaxies through 
conventional optical/NIR spectroscopy because of their faintness in optical/NIR, 
and only two have been confirmed at $z=3.097$ and $z=5.113$, respectively \citep[][]{Wang2019}. 
Blind searches of CO molecular emission line(s) are promising but robust redshift {confirmations are 
limited to a few bright sources with $S_{\rm 850\mu m}>10$ mJy, such as GN10 \citep[$z=5.303$,][]{Riechers2020} 
and HDF850.1 \citep[$z=5.183$,][]{Walter2012}. 
Until recently, precise redshift of the fainter optically dark galaxies has been determined using the blind CO scans with ALMA, 
e.g., ADF22.A2 at $z=3.99$ \citep{Umehata2020}. 
\citet[][]{Zhou2020} performed ALMA spectroscopic scan observations of five fainter sources with $S_{\rm 1.1mm}=1-2$ mJy, 
and suggested two of them likely to have a redshift at $z\sim3.5$.  
The observation of {\sc [C ii]} line is also powerful in confirming the redshift of fainter optically dark galaxies \citep{Schreiber2018}. 
}
All these observations point to redshifts at $z>3$ for optically dark galaxies, consistent with their 
photometric redshifts from UV-to-NIR SED fittings. 
Thanks to its wide spectral coverage of $\sim$7.5 GHz, ALMA is able to constrain redshifts 
through the blind detection of bright emission lines, such as {\sc [C ii]} \citep[e.g.,][]{Swinbank2012}. 
We examined the ALMA datacubes to search for serendipitous emission lines, and found only one possible 
line detection in M0744-H1 above a signal-to-noise ratio (S/N) of 4. As shown in Figure \ref{cii}, the line is detected at $\nu_{\rm obs}=338.609$ GHz 
with a S/N = 4.8. 
{We measured the integrated line flux in the {\tt UV} plane with a point source model for which the central position 
was allowed to vary, and obtained a flux of $0.83\pm0.17$ Jy km s$^{-1}$. 
The bottom-left panel in Figure \ref{cii} shows the continuum-subtracted, velocity-integrated map in the observed frequency range 
338.524--338.696 GHz where the emission line is significant. 
The map is centered at the position of the IRAC source at 3.6\ums (RA=07:44:54.41, DEC=+39:26:45.30). 
It can seen that there is little positional offset between the IRAC source and line map ($<$0$\farcs$15), 
suggesting that both are likely from the same object. 
}
If the line is real, we argue that it could be explained as {\sc [C ii]} emission, as it is the 
brightest emission line within the interstellar medium (ISM) of dusty star-forming galaxies at high redshifts. 
The observation frequency would then suggest a redshift $z=4.613$. Although one spectral line 
alone is not sufficient for a definitive spectroscopic confirmation, the redshift at $z=4.613$ 
is favored as it agrees well with the photometric redshift $z_{\rm phot}=4.66^{+0.54}_{-0.44}$ within errors. 
{The ALMA Band-7 spectra for other galaxies which have brighter flux of $S_{\rm 870\mu m}>$1 mJy are shown in 
the Figure 13 of Appendix C. }

\begin{figure}[htbp]
\centering
\includegraphics[scale=0.38, angle=-90]{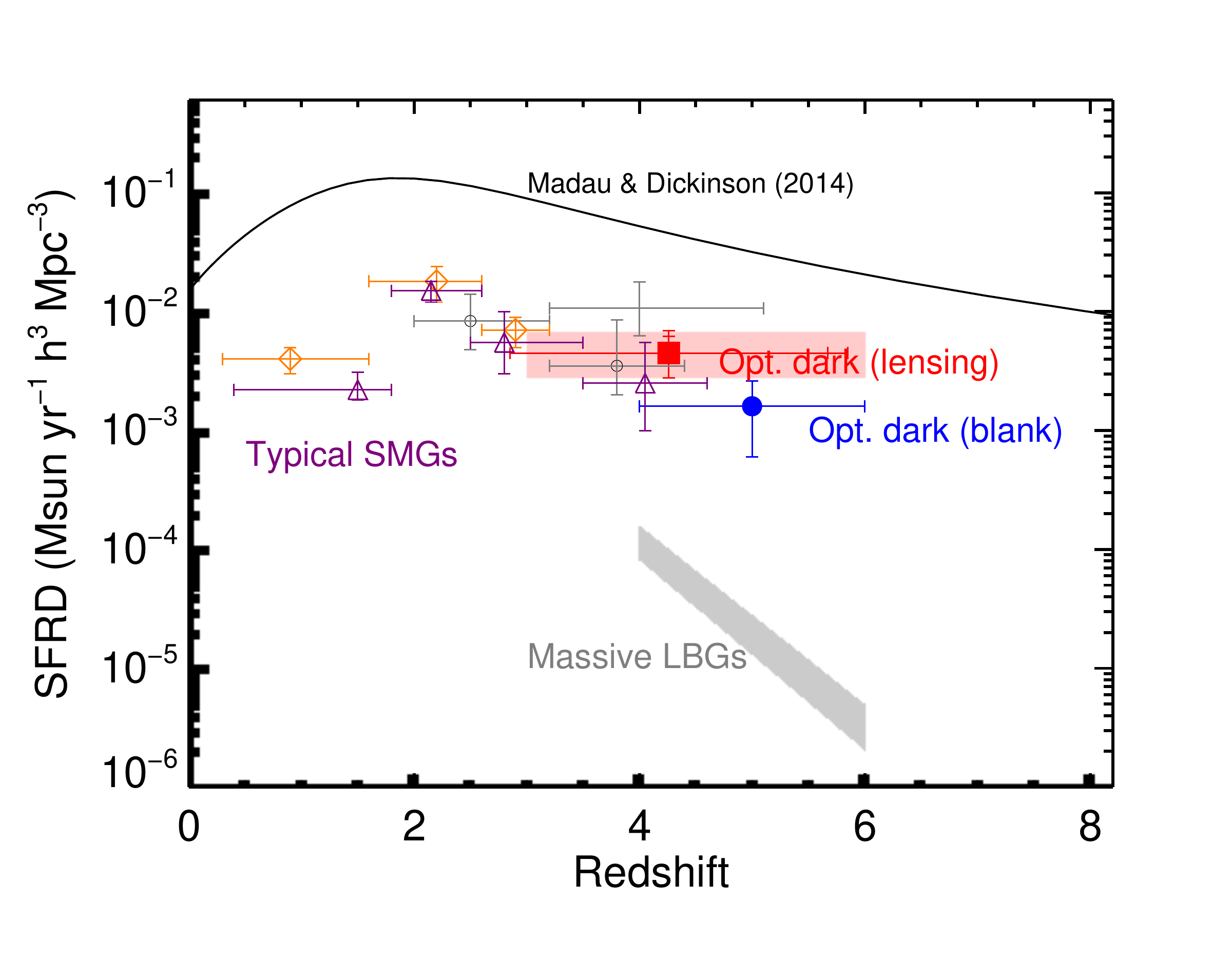}
\vspace{-0.5cm}
\caption{
Estimated contribution of optically dark galaxies in the cluster fields (red filled square) 
to the cosmic SFRD. 
The error bars on SFRD are a combination of 1$\sigma$ Poissonian error on the number of sources and 
the typical IR luminosity error of $\sim$0.15 dex. The black curve shows the UV-based SFRD from a
compilation by \citet[][]{Madau2014}. The literature SFRD values for comparison are from
850\ums selected SMGs \citep[blue diamonds,][]{Chapman2005}, 870\ums selected SMGs \citep[purple triangles,][]{Wardlow2011}, 
and Herschel-SPIRE spectroscopically confirmed $2<z<5$ sources \citep[gray circles,][]{Casey2012}. 
The averaged SFRD contributed by optically dark galaxies in the blank fields is shown in blue filled {circle \citet[][]{Wang2019}.} 
The gray shaded region indicates the SFRD for the UV-selected, massive LBGs with $M_{*}>10^{10.3}$\msun, adopted from \citet[][]{Wang2019}. 
}
\label{sfrd}%
\vspace{-0.05cm}
\end{figure}

\begin{figure*}[htbp]
\begin{center}
\includegraphics[scale=0.65]{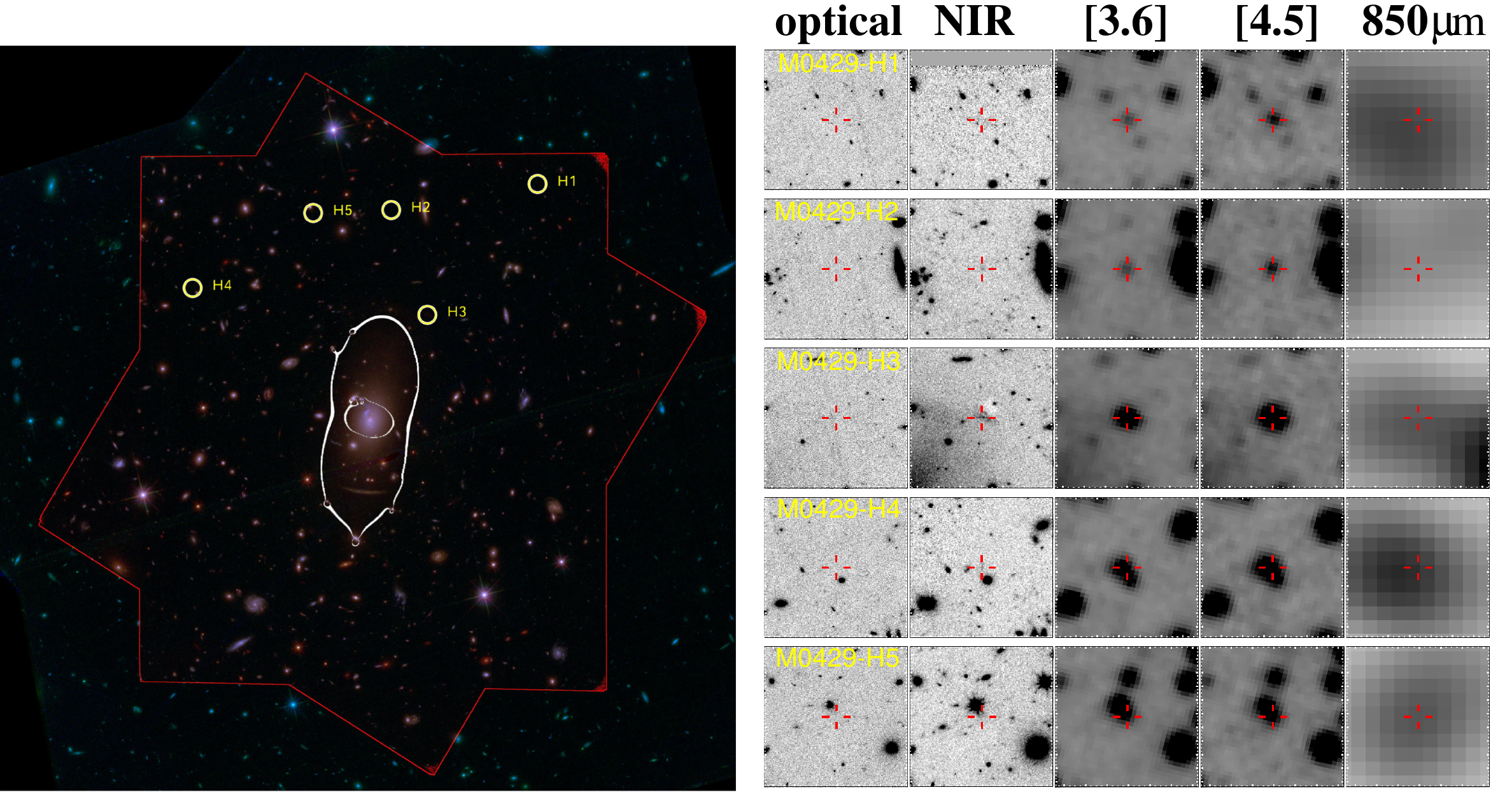}
\vspace{0.1cm}
\end{center}
\caption{{
{\it Left panel:} 
Composite colour image of M0429 made from HST data. The field size with red is $\sim2.1\times2.3$ arcmin$^2$. 
The $z=4$ critical curves (magnification factor $\mu>$75) from the CLASH lensing model are overplotted in white. The locations of the $H$-[4.5\um]$>$4 galaxies are marked with yellow circles. 
{\it Right panel:} 
Cut-out images of five optical dark galaxies. From left to right, the optical images from the respective ACS detection images,  the NIR images from the respective WFC3/NIR detection images, IRAC 3.6\ums and 4.5\ums images, and SCUBA2 850\ums images. 
Each panel has a size of 20\arcsec$\times$20\arcsec.
}}
\label{m0429}%
\vspace{0.5cm}
\end{figure*}

\subsection{Star Formation-rate Density Contribution}

Cosmic Star Formation-rate Density (SFRD) is the integrated star formation in galaxies over a given comoving volume. 
The importance of optically dark galaxies 
to the buildup of stellar mass can be determined by comparing their SFRD contribution to other galaxy populations, 
i.e., UV-selected and submm-selected galaxies. 
This is particularly important for understanding the role of dust obscuration at high redshifts, 
and assessing which fraction being missed in previous optical/NIR surveys. 
Previous studies have found the contribution of dust-obscured star-forming galaxies to the SFRD 
peaking at $z\sim2-3$, where they could contribute half of the total SFRD \citep[e.g.,][]{LeFloch2005, Magnelli2013}. 
At $z>3$, their contribution becomes highly uncertain due to the lack of effective counterpart identifications in the optical/NIR bands, 
so the data to constrain SFRD at this epoch are sparse. 
Contradictory results have so far been obtained in literature, with some claiming that the shape of the SFRD 
does not change much in the redshift range $z=1.5-5.5$ \citep[][]{Swinbank2014, Barger2014}, 
while others suggest an increase of SFRD contribution by dust--obscured star-forming galaxies from $z\sim2.5$ 
to $z\sim4$ \citep[][]{Casey2012, Shu2016, Liu2018}.  

To estimate the contribution of optically dark galaxies in the cluster lensing fields to the cosmic SFRD, 
we used the SFRs derived from the twelve sources that are detected by ALMA and SCUBA2 (Table 1 and Section 4.1). 
Since the sample size is limited, we consider only a single redshift bin of $3<z<6$ within which 
all the detected sources are included based on their photometric redshifts.
The SFRD was then measured by dividing the SFR of each galaxy in the redshift bin by the
comoving volume and summing them. 
The comoving volume is calculated by taking into account the effective survey area of 
cluster lensing fields, which is inversely proportional to the magnification \citep[$A\sim\mu^{-1}$,][]{Bradley2014}. 
Corrected for the lensing effect, the effective survey area is $\sim$70.8 arcmin$^2$ in total. 
The contribution of our sample of lensed optically dark galaxies to SFRD 
is shown in Figure \ref{sfrd} as red filled squares ($4.5^{+1.7}_{-2.4}\times10^{-3}$\sfr Mpc$^{-3}$). 
This is an order of magnitude higher than the SFRD contributed by equivalently massive UV-selected galaxies 
($M_{*}>10^{10.3}$\msun, gray region) at similar redshifts. 
The results support the prevalence of the optically dark galaxies in the early Universe, which 
may dominate the SFRD in massive galaxies \citep[][]{Wang2019}, 
hence they are crucial to the cosmic evolution of most massive galaxies. 


\subsection{Source clustering}

\begin{figure}[htbp]
\centering
\includegraphics[scale=0.6]{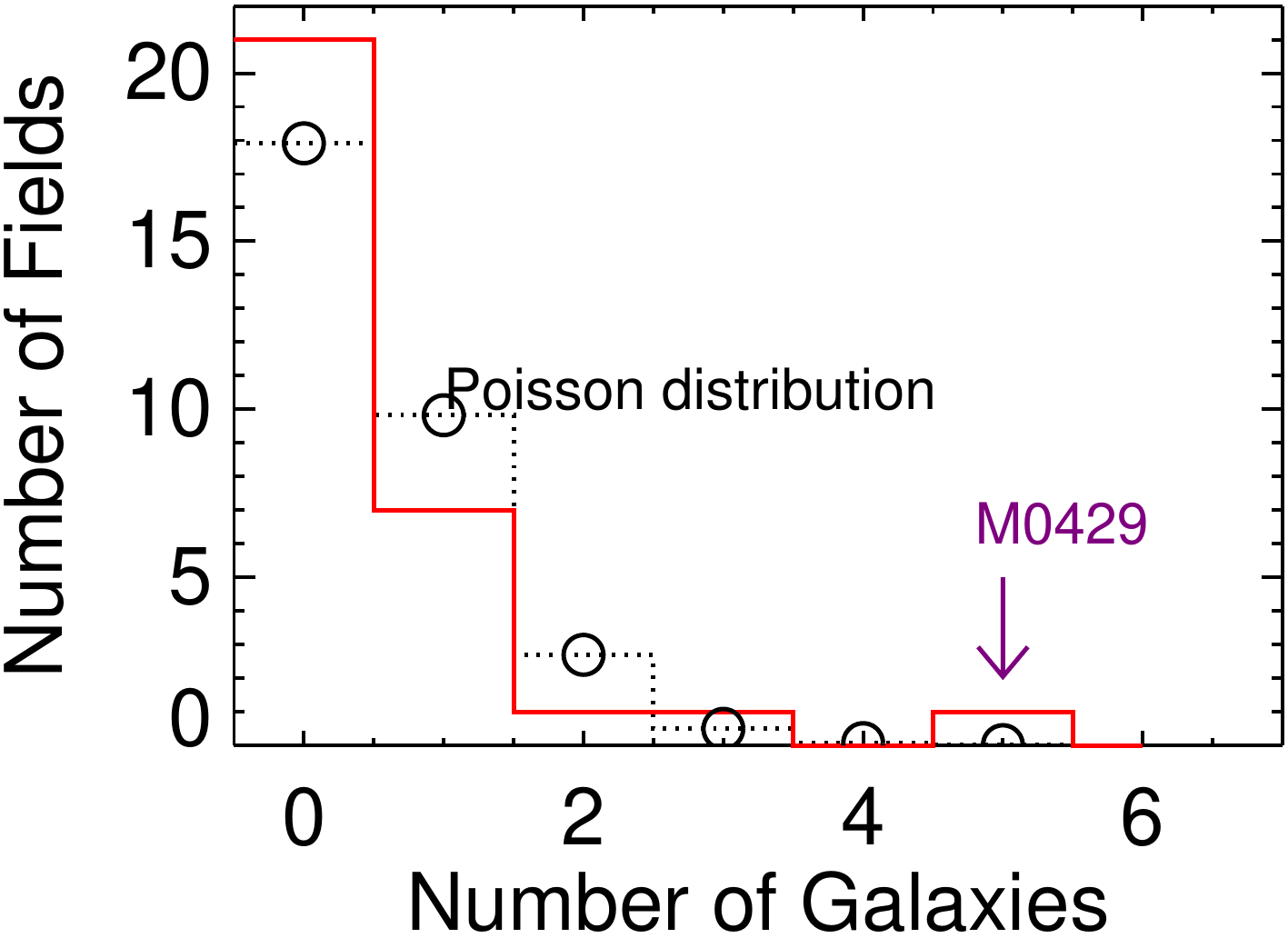}
\caption{
Number-count distribution of the optically dark galaxies within the 31 cluster fields (red solid curve).
Dotted line indicates the number counts expected from the Poisson distribution with the same mean $\langle N \rangle=17/31=0.548$ 
per field. The excess at five in the number of galaxies corresponds to the cluster field of M0429. 
}
\label{poisson}%
\end{figure}

As described in Section 2.3, we identified 17 optically dark galaxies by searching for 31 independent 
{\it HST}/WFC3 pointings of cluster fields\footnote{15 out of 17 optically dark galaxies have been observed 
by ALMA and SCUBA2 (Section 2.4). }. Of them, 21 fields have no sources satisfying our selection criteria 
and seven fields contain a single candidate. Multiple candidates are found in the remaining three fields. 
Figure \ref{m0429} shows the field of M0429 which contains five optically dark galaxies (the highest number among all cluster fields searched). 
Four out of five galaxies are detected by SCUBA2 at 850\ums ($>$3.5$\sigma$), suggesting that they are 
likely dusty star-forming galaxies at high redshifts, consistent with  
their photometric redshift analysis (in the range $z_{\rm phot}=3.8-5.19$).  
Note that unlike arcsec-resolution ALMA data, robust association of 850\ums emission with the counterparts in the optical and NIR bands 
is difficult due to the poor spatial resolution SCUBA2 observations \citep[e.g., ][]{An2018}.   
We quantified the statistical significance of the source clustering in M0429 
by comparing the observed overdensity to the expectation from the null hypothesis (no clustering), 
which is a Poisson distribution with average value of $\langle N \rangle=17/31=0.548$ per field. 
The distribution of the number counts is presented in Figure \ref{poisson}
and there is an apparent excess at $N=5$ (field M0429). 
The remainder of the number-count distribution appears to deviate slightly from Poisson, 
but because of the small number of fields and few number of galaxies per field, 
it is expected that the difference is not statistically significant.
From the simple number-count statistics in the WFC3 field of view, 
we derive that the probability $P(N\ge 5)$ for an overdensity of five dropouts or more 
is 0.00026 under the expectation of $\langle N \rangle=0.548$. 
Therefore, under Poisson statistics, the source overdensity in the field of M0429 is 
significant at $>$99.974\% confidence.

%

Furthermore, all optically dark galaxies in the M0429 field are located within a northern subregion of the $\sim$150\arcsec$\times50$\arcsec~ 
WFC3 field, with a radius of $r<50$\arcsec, suggesting an even more significant source overdensity.  
As such a source overdensity is quite unique among the survey, we performed several checks to demonstrate  
that they are real optical/NIR dark galaxies and represent the missing population of massive galaxies 
at high redshifts. 
All sources are characterized by an extremely red color of $H$-[4.5\um]$>$4 (with a median of 4.3), which would be resulted from 
a strong, redshifted Balmer/4000\AA~break at $z>3$, as indicated by their photometric redshifts. 
In addition, their intrinsic (lensing-corrected) $H$-band magnitudes are in the range $27.1-28.1$ mag, with a median of 27.7 mag, 
well below the detection limit of most {\it HST}/WFC3 observations in blank fields \citep[][]{Wang2019}.  
One source, M0429-H3, has the brightest apparent $H$-band magnitude of 25.46 among the sample. 
{However, it is close to the critical curve of the $z=4$ magnification map (see Figure \ref{m0429}) 
and has a magnification factor as high as 7.86, indicating an intrinsic magnitude of $\sim$27.7. }
This makes the source otherwise not detectable in typical {\it HST}/WFC3 observations. 
Due to the poor resolution of IRAC, blending can cause mis-identifications of sources that are faint in {\it HST}. 
We have addressed this issue by carefully de-blending the IRAC photometry using {\tt GALFIT} (Section 3.2), 
whereby contaminating neighbors are subtracted. 
It has been suggested that a fraction of optically dark galaxies would be missed due to the blending with nearby 
bright optical sources \citep[e.g.,][]{Franco2018, Zhou2020}. 
Our analysis is thus relatively complete in searching for optically dark galaxies.


The comparison with Poisson statistics shows that the overdensity of sources in field M0429 
is physical and not originating from random statistical fluctuations at a confidence of $>$99.974\%. 
{Lensing is unlikely to cause the apparent overdensity, i.e., multiple distorted images of the same 
background source, because the median magnification in that area is not high ($\mu=$2.4). 
We checked for the counter images predicted by the lensing models (Section 3.1) for the source 
M0429-H3 that has the largest magnification factor, but did not find its lensed counterparts. }
\citet[][]{Wang2019} found that the optically dark galaxies in CANDELS fields appear to be strongly clustered, 
as measured from their angular cross-correlation function with UV-selected galaxies at $3.5<z<5.5$ in the same fields.  
Given the relatively small-area of \hst~observations of cluster fields, a fluctuation in the number counts 
due to cosmic variance should be taken into account. 
By assuming a mean number density of $\langle N \rangle$ = 0.548 optically dark galaxy per WFC3 field 
and galaxy bias $b$ =8.4 \citep[][]{Wang2019}, 
we derived a cosmic variance of 47\% from the cosmic variance calculator \citep[][]{Trenti2008}\footnote{https://www.ph.unimelb.edu.au/~mtrenti/cvc/CosmicVariance.html}. 
This means that the overdensity of five optically dark galaxies in M0429 is still significant at 7.3$\sigma$ 
after considering the effects of cosmic variance.  
It is probable that the overdensity in M0429 is attributed to the projected galaxy clustering 
from sources that are at different redshifts having similarly red $H-$[4.5\um] colors. 
Alternatively, the overdensity structure could be associated with a protocluster of galaxies 
falling into a narrow redshift range. 
For example, \citet[][]{Zhou2020} found that four out of six optically dark galaxies in the GOODS-ALMA field 
reside in a small area of 5 arcmin$^2$, and their redshift distribution concentrates at $z\sim$3.5, suggesting 
an association with a proto-cluster structure. 
Note that overdensties in the high-redshift UV-selected galaxies ($z>7$) in the cluster fields have been previously 
reported \citep[][]{Trenti2012, Zheng2014, Ishigaki2016}, but our findings of five optically dark galaxies within $\sim$50\arcsec~($\sim$200 kpc in the source plane) is unique. 
Future spectroscopy observations with ALMA and/or JWST will be helpful to constrain the nature of the overdensity in M0429, 
and test whether these galaxies can trace a massive galaxy cluster in formation.


\section{Conclusions}

We have presented ALMA 870\ums observations of ten optically dark galaxies in the cluster lensing fields. 
We detect the dust emission in eight out of ten galaxies down to $\sim$0.6 mJy. Supplementary to the ALMA 
sample, we observed five optically dark galaxies using JCMT/SCUBA2, from which four are detected at 850\ums 
down to 1.6 mJy. This makes this work one of most sensitive searches for dust emission in optically dark galaxies. 
Thanks to gravitational lensing, we reach a deeper limiting flux (by a factor of $\sim$3) 
in comparison with blank fields, confirming the presence of a population of intrinsically faint, 
massive star-forming galaxies at high redshifts that could be missed in deep optical/NIR surveys even with {\it HST}. 
Five galaxies have bright MIR emission with 4.5\ums magnitude of less than 22, which 
are suitable for follow-up spectroscopy observations, e.g., with JWST.    
After corrected for lensing effect, the SFR density of  optically dark galaxies in cluster fields is estimated 
to be $4.5^{+1.7}_{-2.4}\times10^{-3}$\sfr Mpc$^{-3}$, which is an order of magnitude higher than that of 
equally massive UV-selected galaxies at similar redshifts, consistent with the result obtained in blank fields. 
Significant source clustering is observed on the physical scale of $\sim$200 kpc. 
Given the large errors on photometric redshifts, we cannot tell 
whether the overdensity is due to the projected galaxy clustering from sources 
at different redshifts, or associated with a protocluster of galaxies. 
The latter scenario, if confirmed with follow-up spectroscopic observations, will have important 
implications for the formation and evolution of most massive galaxies in the early Universe. 
\\
\\
\bigskip
 \begin{acknowledgments}
 The data presented in this paper are based on observations
 made with the NASA/ESA Hubble Space Telescope from the CLASH Multi-Cycle Treasury Program 
 (GO-12065) and the Frontier Fields program conducted by STScI, which is operated by the Association 
 of Universities for Research in Astronomy,
Inc. under NASA contract NAS 5-26555. 
The work is also based on data obtained with the Spitzer Space Telescope, which is operated
by the Jet Propulsion Laboratory, California Institute of Technology under a contract with NASA.
 The authors thank ALMA and JCMT operations staff for their assistance in scheduling and performing the observations. 
 This paper makes use of the following ALMA and JCMT data: ALMA \#2018.1.01409.S and JCMT \#M16BP006. 
 ALMA is a partnership of ESO (representing
 its member states), NSF (USA) and NINS (Japan), together with NRC
 (Canada), MOST and ASIAA (Taiwan), and KASI (Republic of Korea), in cooperation
 with the Republic of Chile. The Joint ALMA Observatory is operated
 by ESO, AUI/NRAO and NAOJ. 
 The James Clerk Maxwell Telescope is operated by the East
 Asian Observatory on behalf of The National Astronomical
 Observatory of Japan, Academia Sinica Institute of Astronomy
 and Astrophysics, the Korea Astronomy and Space Science
 Institute, the National Astronomical Observatories of China,
 and the Chinese Academy of Sciences (Grant No.
 XDB09000000), with additional funding support from the
 Science and Technology Facilities Council of the United
 Kingdom and participating universities in the United Kingdom
 and Canada.
 The work is supported by Chinese NSF through grant No. 11822301. 
 W.H.W. acknowledges the grant support from the Ministry of Science and Technology
 of Taiwan (110-2112-M-001-006). 
 Y.H. acknowledges support from Chinese NSF through grant No. 11773063, and 
 Natural Science Foundation of Yunnan Province (grant No. 2017FB007). 

\end{acknowledgments}

\software{ SEXTRACTOR (Bertin \& Arnouts 1996), CASA (v4.3.1; McMullin et al. 2007), SMURF (Chapin et al. 2013), Starlink (Currie et al 2014), GALFIT software (Peng et al. 2010), EAZY (Brammer et al. 2008), BayeSED (Han \& Han 2012, 2014, 2019), FAST (Kriek et al. 2009)}

\begin{deluxetable*}{cccccccccccc}[htbp]
\tablenum{1}
\tablecaption{Photometry and stellar properties of the optically dark galaxies in cluster fields}
\tablewidth{0pt}
\tablehead{
\colhead{Name} &
\colhead{R.A.} &
\colhead{Decl.} &
\colhead{F160W} &
\colhead{3.6\ums} &
\colhead{4.5\ums} &
\colhead{870\ums} &
\colhead{$\mu$} & 
\colhead{$z_{\rm photo}$} &
\colhead{Log$M_{\star}$} \\
 & (J2000) & (J2000) & mag & mag & mag & (mJy) &  &  & \msun \\
 (1) & (2) &(3) &  (4) & (5)  & (6) & (7) & (8) & (9)  & (10) \\
}
\startdata
A2744-H1 & 00:14:16.025 & -30:23:41.42 & 26.08\ppm0.11  &22.18\ppm0.08  &  21.64\ppm0.04  & 1.245\ppm0.065   &  1.95   &  4.11$^{+0.13}_{-0.13}$  & 11.49$^{+0.02}_{-0.04}$ \\
Bullet-H1 & 06:58:39.79 & -55:56:57.77 & 25.70\ppm0.08$^\dag$  &  21.40\ppm0.08 &   20.79\ppm0.08 &  6.7\ppm0.19    &  $>$8  &  5.02$^{+0.39}_{-0.36}$ &  11.68$^{+0.37}_{-0.13}$ \\
M0744-H1 & 07:44:54.44 & +39:26:45.15  &  26.69\ppm0.36$^\dag$   & 22.87\ppm0.12 & 22.05\ppm0.06  & 1.42\ppm0.15   &  2.53 &  4.66$^{+0.54}_{-0.44}$ & 11.65$^{+0.44}_{-0.68}$\\
M0744-H2 & 07:44:50.91 & +39:26:13.5 & 26.38\ppm0.41$^\dag$              & 24.36\ppm0.36   &  24.18\ppm0.31  &  $<$0.140    &   1.42 &  2.66$^{+1.1}_{-1.0}$ & 9.04$^{+1.2}_{-0.36}$  \\
M0744-H3$^{\S}$ &07:44:51.05  & +39:28:43.5  &    $\dots$           & 22.88\ppm0.16   &  22.18\ppm0.09  &  $<$0.172   &  1.58 &     $\dots$ & $\dots$ \\
M1115-H1 & 11:15:53.60 & +01:30:46.7 & 26.00\ppm0.17$^\dag$  &  23.20\ppm0.15  &  22.68\ppm0.11  & 0.617\ppm0.041 &  1.86 &  3.68$^{+0.62}_{-0.55}$ & 10.28$^{+0.80}_{-0.32}$\\
M1115-H2 & 11:15:54.28 & +01:30:36.4 & 25.74\ppm0.08      &  22.93\ppm0.13  &  22.03\ppm0.15   &  2.79\ppm0.28   & 1.91 & 3.44$^{+0.21}_{-0.23}$ & 10.93$^{+0.12}_{-0.70}$ \\  
A2261-H1 & 17:22:31.92 & +32:07:53.8  &  25.95\ppm0.09  &  22.23\ppm0.10  &  21.79\ppm0.11  & 3.41\ppm0.11   &     1.71   &    4.37$^{+0.30}_{-0.28}$ & 11.4$^{+0.2}_{-0.03}$ \\
M2129-H1 & 21:29:27.04 & -07:42:50.8 &  $>$27.2$^\dag$        & 24.42\ppm0.25  &  23.77\ppm0.23   &  2.2\ppm0.044   & 1.4 &  4.41$^{+0.71}_{-0.67}$ & 10.90$^{+0.65}_{-0.77}$ \\
M2137-H1 & 21:40:21.19 & -23:39:06.8 &  25.79\ppm0.39$^\dag$    & 22.60\ppm0.15  &  22.07\ppm0.15  &  0.586\ppm0.056 & 1.37 &  3.95$^{+0.51}_{-0.51}$ & 11.18$^{+0.37}_{-0.65}$ \\ 
M0429-H1 & 04:29:33.05 & -02:52:04.8 & 27.28\ppm0.20   &  23.11\ppm0.25   & 22.50\ppm0.13 &  2.3\ppm0.45$^\ddag$   &  1.76 & 4.74$^{+0.28}_{-0.27}$ & 11.38$^{+0.04}_{-0.16}$\\
M0429-H2 &  04:29:35.64 & -02:52:11.6 & 27.02\ppm0.13  &    23.16\ppm0.26  & 22.72\ppm0.18 & $<1.5$$^\ddag$       &  2.8  &  4.26$^{+0.22}_{-0.27}$ & 11.07$^{+0.06}_{-0.16}$\\
M0429-H3 & 04:29:35.00 & -02:52:39.0 & 25.46\ppm0.06    &  22.61\ppm0.05   & 21.24\ppm0.03  & 1.61\ppm0.42$^\ddag$  &      7.86 & 3.80$^{+0.23}_{-0.19}$ & 11.01$^{+0.27}_{-0.13}$\\
M0429-H4 & 04:29:39.15 & -02:52:32.0  &26.40\ppm0.11    &  21.82\ppm0.06   & 21.13\ppm0.03  & 6.9\ppm0.45$^\ddag$   &     1.92 & 5.19$^{+0.40}_{-0.33}$ & 12.27$^{+0.05}_{-0.32}$\\
M0429-H5 & 04:29:37.02 & -02:52:12.35  &  26.19\ppm0.18$^\dag$         &  22.91\ppm0.13   & 22.14\ppm0.11   &  4.7\ppm0.44$^\ddag$  &  2.42   &  4.13$^{+0.58}_{-0.55}$ & 10.73$^{+0.85}_{-0.33}$\\
\enddata
\caption{
Columns (1):  Source ID; (2) \& (3) Coordinate. For sources detected by ALMA, the centroid of submm emission is adopted, otherwise the coordinate 
represents the centroid of IRAC 4.5\ums emission; 
(4) \hst~WFC3/F160W AB magnitude. $^\dag$Sources are not shown in the public CLASH/HFF catalog, for which manual \hst~photometry centered 
on the position of ALMA source is performed. 3$\sigma$ limiting magnitude is given for non-detection. 
$^{\S}$\hst~images for this source are contaminated by stellar diffraction spikes, which are difficult to remove for reliable photometry. 
(5) \& (6) IRAC 3.6\ums and 4.5\ums photometry; (7) Integrated 870\ums flux observed by ALMA. $^\ddag$For the cluster M0429, the 850\ums flux (or upper limit) observed by JCMT/SCUBA2 is reported; (8) Magnification factor; (9) Photometric redshifts are derived 
from EAZY with 1$\sigma$ errors; (10) Stellar masses are derived from FAST with 1$\sigma$ errors. Note that all the photometry and physical parameters are observed ones, 
i.e., not corrected for the lensing effects. 
}
\end{deluxetable*}
\clearpage
\bibliography{ms_hdrop_almacluster.bbl}
\end{sloppypar}

\appendix
\section{SED fitting results}
\begin{figure*}[htbp]
\centering
\includegraphics[scale=0.2]{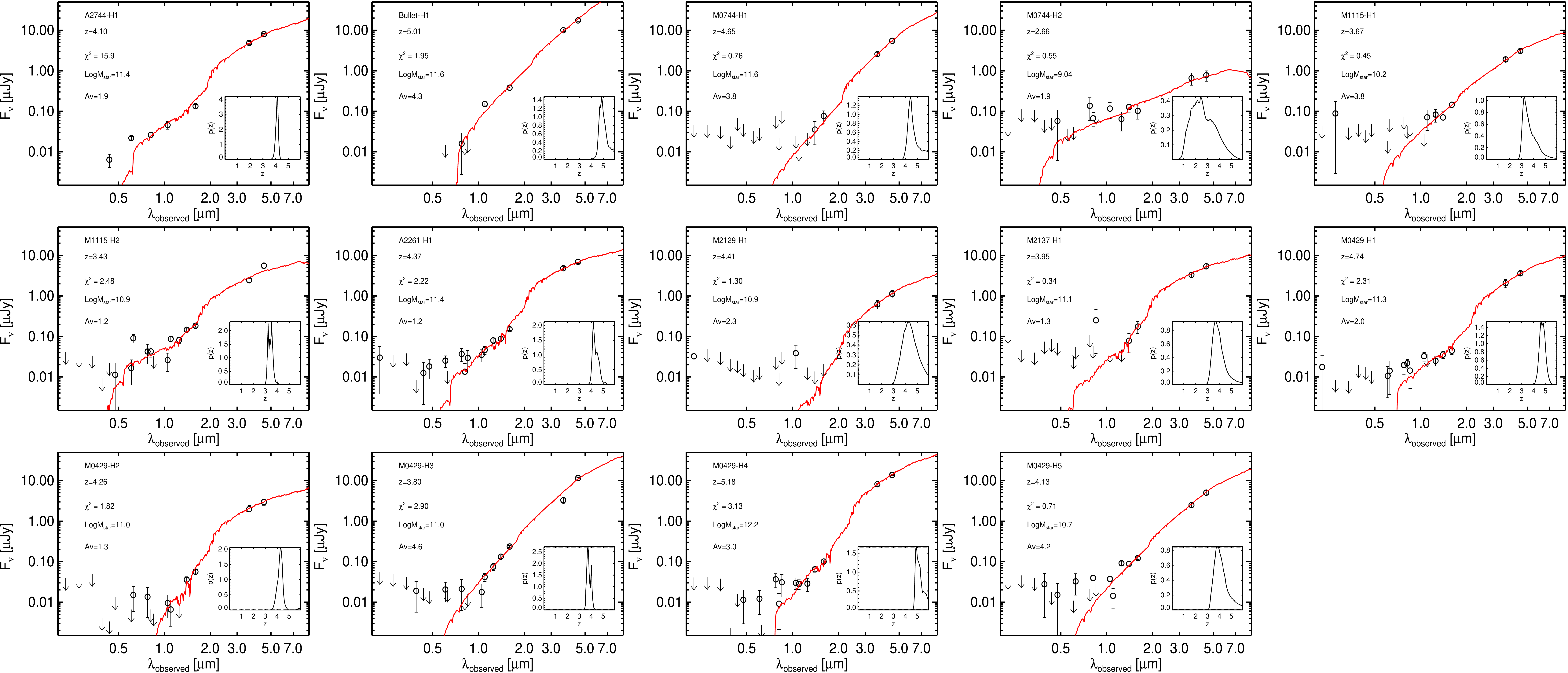}
\label{fastsed}%
\caption{
The observed SED and its best-fit model from {\tt FAST} for galaxies listed in Table 1. Inset panel shows the probability distribution of photometric redshift from {\tt EAZY}. 
 }
\end{figure*}

\section{Tests on the accuracy of photo-z analysis and SED fittings}

{Since the galaxies in our sample, by selection, have extremely red $H$-[4.5\um] colors, 
the standard photo-z analysis and SED fittings may suffer from the uncertainty in the 
stellar modeling and stellar parameters, such as the degeneracies between 
redshift, dust extinction and stellar masses. 
In addition, there may have the effect of systematics from the choice of specific assumptions and 
priors in the methods adopted. 
To understand these uncertainties, we use the Bayesian SED modeling code--BayeSED \citep[][]{Han2012, Han2014, Han2019} 
to self-consistently estimate 
the photometric redshift and infer the physical properties for each galaxy. 
By employing efficient machine learning methods, such as the artificial neural network (ANN) 
algorithm and the Bayesian inference tool--MultiNest, 
BayeSED enables a continuous sampling of the parameter space and provides the posterior 
probability distribution function (PDF) of all modeling parameters. This allows us to recognise 
the degeneracies between parameters and assess the parameter uncertainties. 
We choose the same stellar populations library as used in FAST (Section 3.2). 
Figure 10 (a) shows the comparison of photo-$z$ between BayeSED and EAZY. 
While the redshift distribution given by BayeSED is slightly lower than EAZY, the median redshift for the sample 
($z_{\rm median}=3.39$) is close to that obtained by EAZY within errors (Section 3.2). 
Except for one source, the stellar masses are consistent with each other, as shown in Figure 10 (b). 

Based on the results derived from BayeSED, we further tested whether the photo-$z$ analysis and stellar properties can be 
better constrained by using the infrared luminosities ($L_{\rm IR}$) inferred from the ALMA/SCUBA-2 fluxes (or upper limits for non-detections). 
Firstly, we used the infrared luminosities extrapolated from the ALMA/SCUBA-2 fluxes by assuming that the intrinsic FIR SED resembles 
that derived from the stacked Herschel fluxes \citep{Wang2019}. 
The observed infrared luminosity for each galaxy was then compared to the model value to put constraint on the dust attenuation. 
The resulting distribution in photo-$z$ by applying the $L_{\rm IR}$ constraints is slightly lower than that without constraints, 
with a median redshift of $z_{\rm median}=3.08$ (Figure 10 (c)). The effect on the stellar masses can also be seen in Figure 10 (d), 
which indicates a difference of 0.1 dex according to the median value in the distribution. 
In Figure 11, we present the 1D and 2D posterior PDFs of relevant model parameters for BayeSED (blue) and 
that by applying the $L_{\rm IR}$ constraints (red). 
For comparison, we also present the results from the $L_{\rm IR}$ constraints by assuming that the intrinsic FIR SED can be described 
by a greybody model with a characterized dust temperature of 36.7 K \citep{Wang2019} to extrapolate infrared luminosities (grey).  
As shown in the figure, applications of the $L_{\rm IR}$ constraints can lead to somewhat different shapes of posterior PDFs. 
However, most model parameters are still not well constrained as the posterior PDF is broad. 
Although the shape of posterior PDF appears to be slightly different, the parameter estimation derived from 
the median of the posterior PDFs is actually not deviating too much in most cases. 
Therefore, we conclude that applying the $L_{\rm IR}$ constraints in the SED fittings can lead to smaller values 
of photo-$z$, likely due to the constraints on dust attenuation, while the effect on the stellar masses is minor. 
Note that the results do not change if the delayed-$\tau$ model (instead of direct-$\tau$) is assumed for the parameterization of the SFH, 
as shown in Figure 12. 
On the other hand, high-resolution submillimeter observations have shown that some ALMA-detected sources display significant spatial offsets 
between the positions of dust emission and the UV/optical emission \citep[e.g.,][]{Elbaz2018, Franco2020}. 
Due to the faintness of the rest-frame UV emission observed by \hst, this can not be tested with the current data. Therefore, 
considering that the dust continuum emission may not completely be the one obscuring the stellar component, in this paper 
we report only the results of photo-$z$ analysis and SED fittings without applying the $L_{\rm IR}$ constraints. 



}

\begin{figure*}[htbp]
\centering
\includegraphics[scale=0.2]{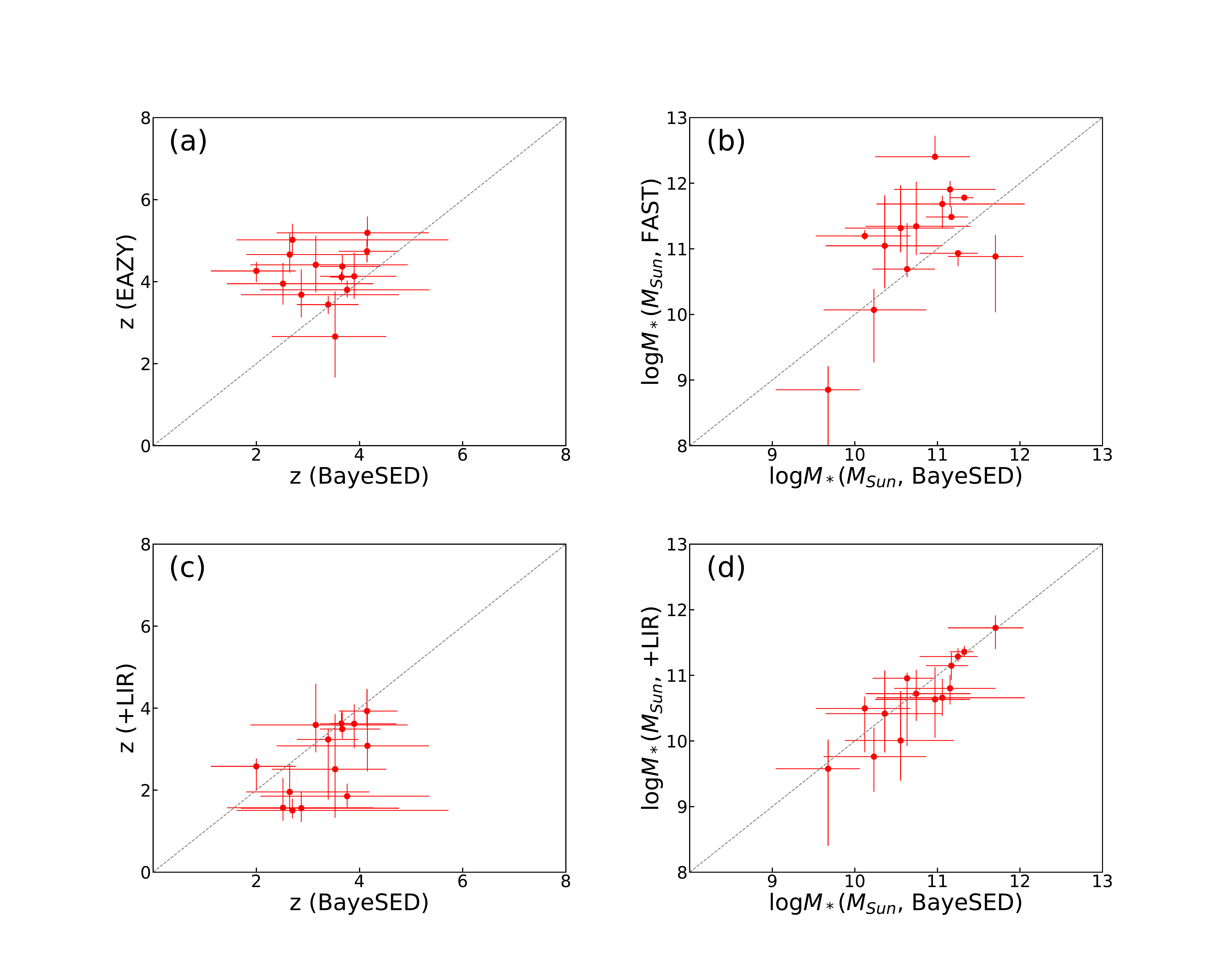}
\label{zcompa}%
\caption{
Comparison of photo-$z$ derived from the SED fitting code EAZY with that of BayeSED (panel (a)). 
The corresponding comparison of stellar masses derived from FAST with that of BayeSED is shown in panel (b). 
Panel (c) and (d) shows the comparison of photo-$z$ and stellar masses derived by BayeSED 
with and without $L_{\rm IR}$ priors. 
The dashed line in each panel represents the one-to-one relation. 
 }
\end{figure*}

\begin{figure*}[htbp]
\centering
\includegraphics[scale=0.45]{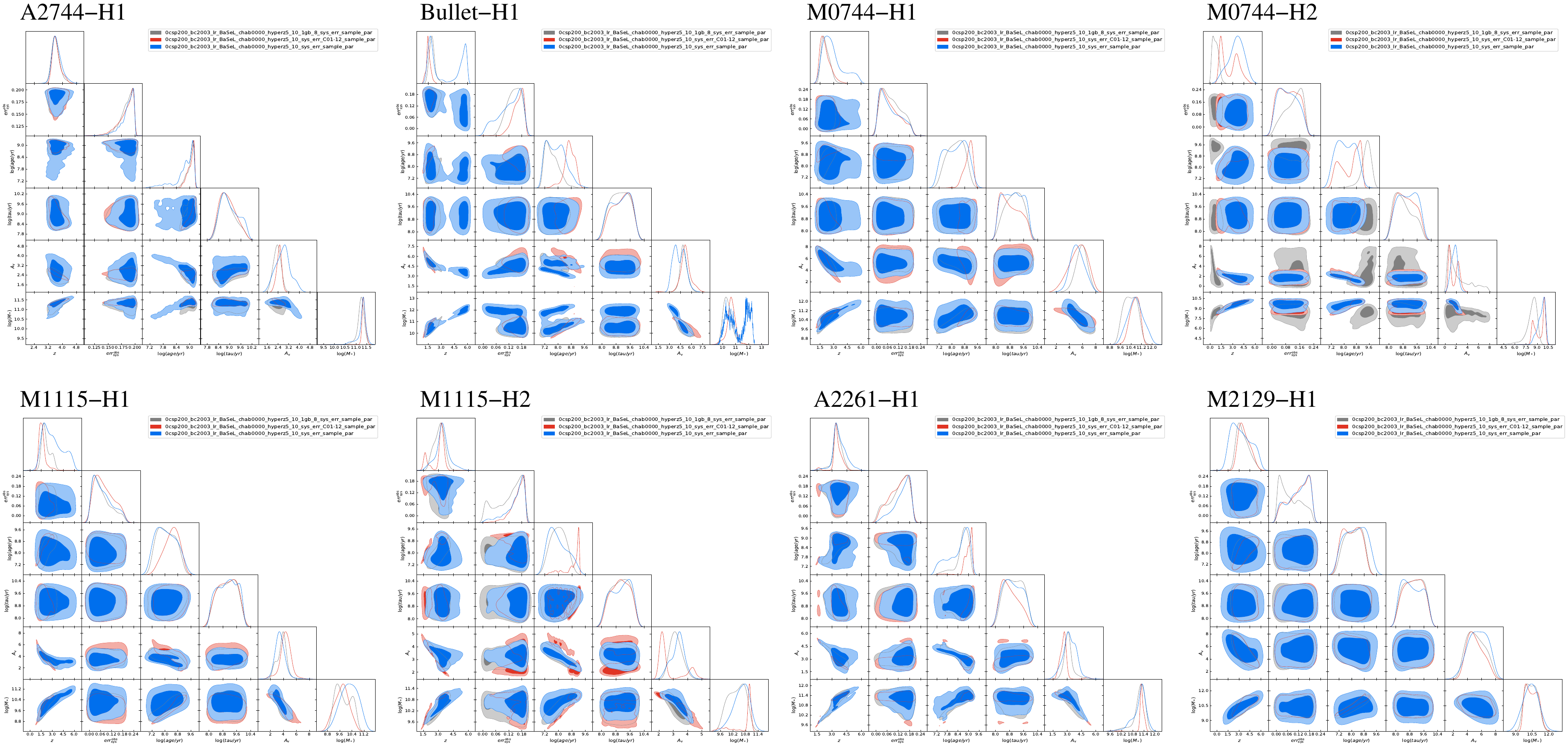}
\label{zcompa}%
\caption{
1D and 2D posterior PDFs of relevant model parameters for BayeSED (blue) and 
that by applying the $L_{\rm IR}$ constraints (red). 
As a comparison, the results from the $L_{\rm IR}$ constraints by assuming the intrinsic FIR SED with a greybody model are 
also presented (grey).  
 }
\end{figure*}

\setcounter{figure}{10}

\begin{figure*}[htbp]
\centering
\includegraphics[scale=0.45]{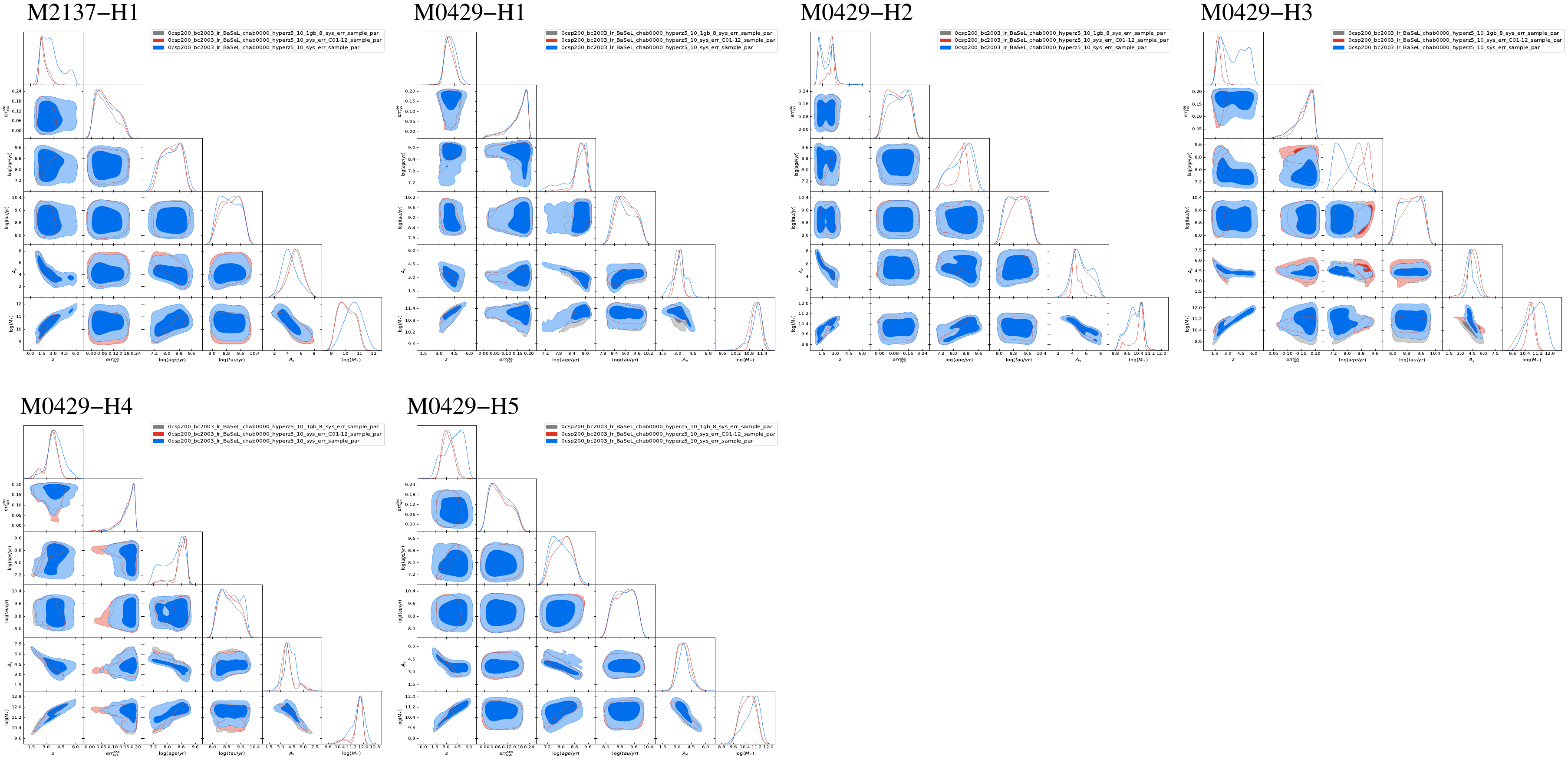}
\label{zcompa}%
\caption{
 {Continued.}
 }
\end{figure*}

\begin{figure*}[htbp]
\centering
\includegraphics[scale=0.2]{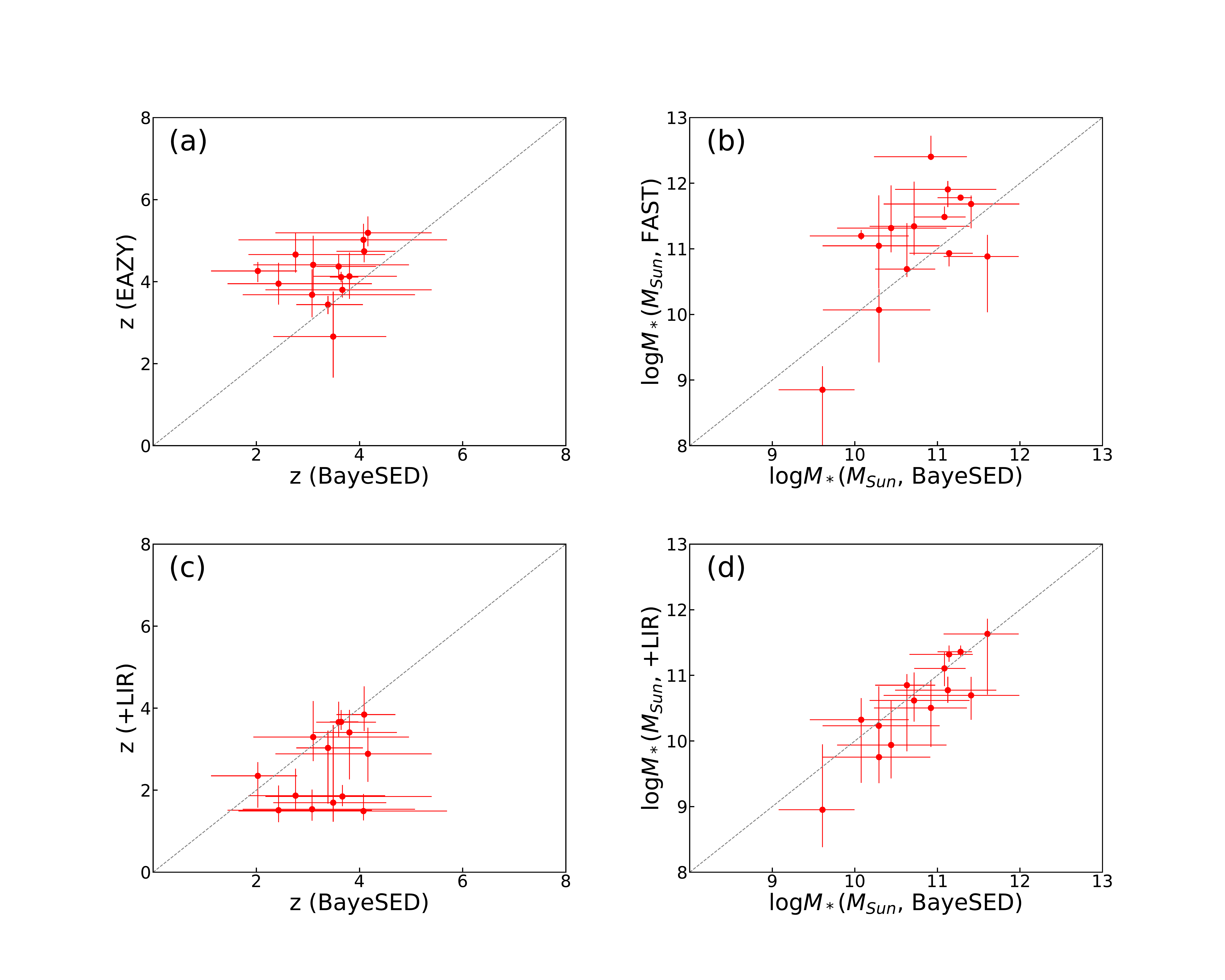}
\label{zcompa}%
\caption{
The same as Figure 10, but assuming delayed-$\tau$ model (instead of direct-$\tau$) for the parameterization of the SFH. 
 }
\end{figure*}

\clearpage

\section{ALMA Band-7 spectrum for galaxies with flux of $S_{870\mu\lowercase{m}}>1$\lowercase{m}J\lowercase{y}}

\begin{figure*}[htbp]
\centering
\includegraphics[scale=0.6]{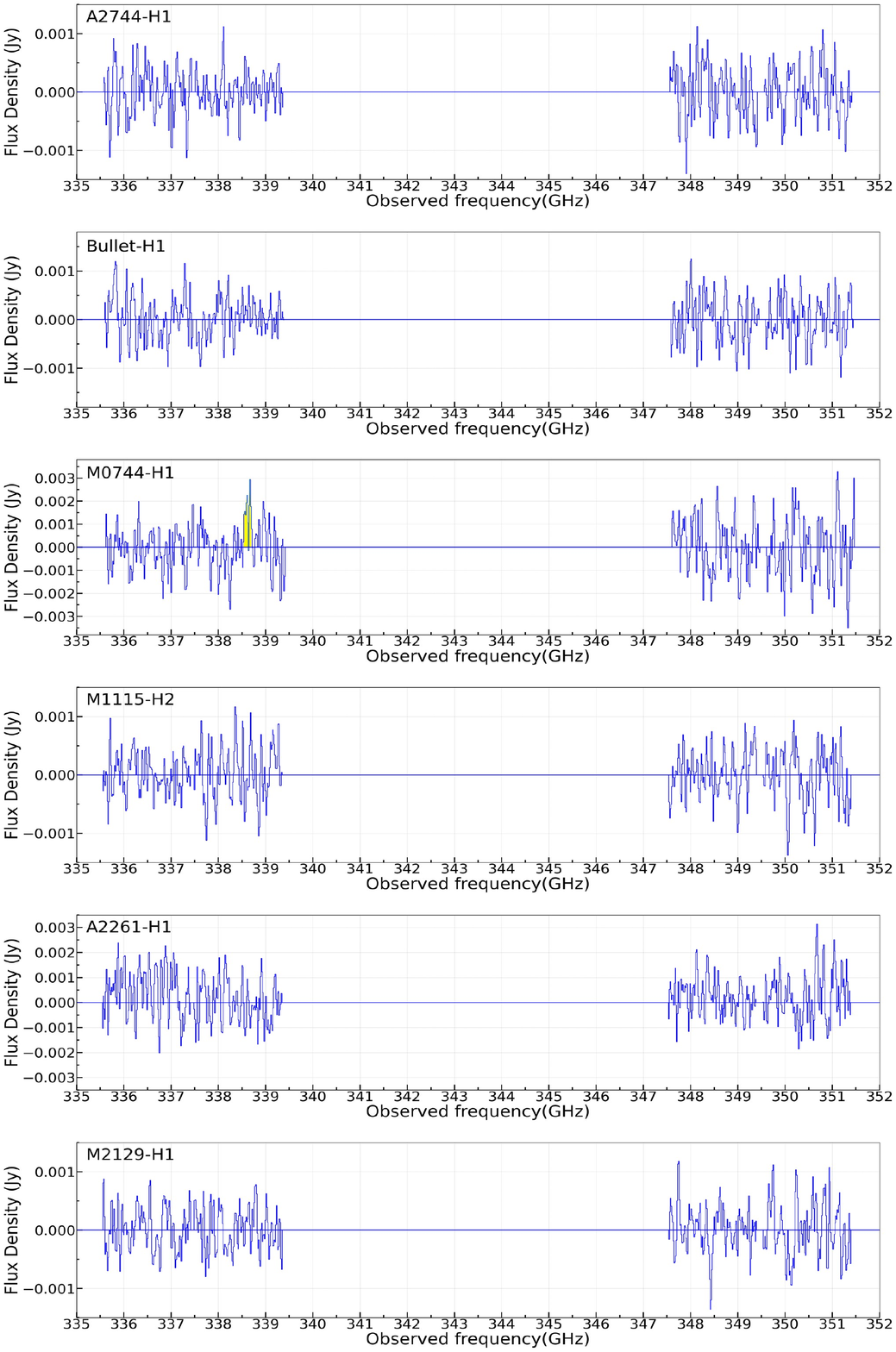}
\label{zcompa}%
\caption{
ALMA Band-7 spectrum for galaxies with brighter flux of $S_{870\mu\lowercase{m}}>1$\lowercase{m}J\lowercase{y},  
which covers the frequency range 335.5-339.5 GHz and 347.5-351.5 GHz. Except for M0744-H1, no obvious emission 
lines above $4\sigma$ are detected in other galaxies.  
 }
\end{figure*}

\end{document}